\documentclass[aps,twocolumn,groupedaddress,pra]{revtex4-1}
\usepackage{graphicx}
\usepackage{mathrsfs}
 \usepackage{amsmath,amssymb,amsfonts}

\newcommand{\be}{\begin{eqnarray}}
\newcommand{\ee}{\end{eqnarray}}

\begin{document}


\title{Decoherence of trapped bosons by buffer gas scattering: What length scales matter?}


\author{Lukas Gilz}
\email[]{lgilz@rhrk.uni-kl.de}
\author{Luis Rico-P\'erez}
\author{James R. Anglin}

\affiliation{\mbox{State Research Center OPTIMAS and Fachbereich Physik,} \mbox{Technische Univerit\"at Kaiserslautern,} \mbox{D-67663 Kaiserslautern, Germany}}


\date{\today}

\begin{abstract}
We ask and answer a basic question about the length scales involved in quantum decoherence: how far apart in space do two parts of a quantum system have to be, before a common quantum environment decoheres them as if they were entirely separate? We frame this question specifically in a cold atom context. How far apart do two populations of bosons have to be, before an environment of thermal atoms of a different species (`buffer gas') responds to their two particle numbers separately? An initial guess for this length scale is the thermal coherence length of the buffer gas; we show that a standard Born-Markov treatment partially supports this guess, but predicts only inverse-square saturation of decoherence rates with distance, and not the much more abrupt Gaussian behavior of the buffer gas's first-order coherence. We confirm this Born-Markov result with a more rigorous theory, based on an exact solution of a two-scatterer scattering problem, which also extends the result beyond weak scattering. Finally, however, we show that when interactions within the buffer gas reservoir are taken into account, an abrupt saturation of the decoherence rate does occur, exponentially on the length scale of the buffer gas's mean free path. 
\end{abstract}


\maketitle

\section{Introduction}
While dissipation and noise are the basic phenomena of classical open systems \cite{gardiner_handbook_2004}, quantum open systems suffer decoherence as well \cite{Zurek_Decoherence_2003}. This uniquely quantum phenomenon depends on the exchange of information between open system and environment, rather than on the exchange of energy. Decoherence can be induced by system-environment couplings of the same form as those responsible for dissipation and noise, and it can be computed with the same theoretical techniques that describe those more familiar non-equilibrium processes \cite{BL_coherent}. Since decoherence concerns the exchange of information rather than of energy, however, it is in some respects fundamentally different from dissipation and noise. Decoherence can sometimes occur, for example, on very much shorter time scales than dissipation \cite{Zurek_Decoherence_2003}. Decoherence can also be drastically non-classical in regard to length scales, for although interactions between system and environment are always local, quantum states and quantum information are not. Quantum mechanics fulfills the political activist's slogan, ``Act locally, think globally.'' 

In this paper we investigate what length scales matter for decoherence when a large volume of ideal gas in the Maxwell-Boltzmann temperature regime, acting as a reservoir, interacts with an idealized cold atom model whose initial state is a coherent quantum superposition of different distributions of particles between two spatially separated trapping sites. The local interactions between system and reservoir take the form of collisions between the two types of particles. The reservoir gas particles (``buffer gas'') are of a distinct type from the trapped population, so that there is no exchange of particles between system and reservoir; by taking the system particles to be very tightly trapped, we also turn off energy exchange, in order to focus strictly on information exchange and decoherence. This scenario is one that could in principle be realized accurately in today's cold atom laboratories \cite{spethmann_inserting_2012}, but it is also qualitatively representative of more naturally occurring situations in which distant portions of a continuously extended Bose-Einstein condensate suffer decoherence of their relative phase due to interaction with a thermal cloud \cite{graham_decoherence_1998}.

The buffer gas reservoir effectively measures the spatial distribution of the system particles, by colliding with them wherever they are; here we investigate the spatial resolution of this effective measurement. We focus specifically on the way in which decoherence of different number distributions strengthens with increasing distance between the two sites. In the limit where the two traps are so close as to overlap completely, the environmental gas will couple only to their total occupation, and hence induce no decoherence at all between quantum superpositions of the same total number of particles being differently distributed between the two traps. The farther the traps are apart, the better the thermal gas bath can resolve how many particles are held in each one, and so the faster superpositions of different number distributions will decohere. Ultimately, however, one expects that there should exist some saturation distance, beyond which decoherence rates rapidly approach their asymptotic values at infinite separation. See Figure 1.

The subject of our paper is this nontrivial dependence of the decoherence rate on the separation distance across which particles are coherently superposed. In the end we identify two generically important length scales of the reservoir, to which the spatial separation between parts of the open system should be compared: the thermal coherence length and the mean free path. We explain the roles of both scales. For the collisions that we consider between system and buffer gas particles, the s-wave scattering length is another length scale, which determines the strength of local system-reservoir interactions. We obtain a general formal expression for the dependence of decoherence rates on the scattering length as well. In its most general form this expression is complicated, but it simplifies in the limit of weak scattering.

We compute our decoherence rates using a succession of different theoretical techniques. After deducing the importance of the buffer gas coherence length as a mere qualitative estimate based on the equilibrium properties of the reservoir, we derive a Born-Markov master equation, such as is routinely applied to open quantum systems in quantum optics \cite{Gardiner_Quantum_2004, Scully_Quantum_1997, walls_quantum_2007}, condensed matter \cite{Weis_Quantum_1999,dattagupta_dissipative_2004} and cold atomic vapors \cite{Anglin_Cold_1997,Gilz_Quantum_2011, Gardiner_Quantum_1997,Gardiner_Quantum_1998, ruostekoski1998,Schelle_Number_2011}. Our Born-Markov analysis confirms the relevance of the buffer gas coherence length, but predicts a relatively slow, inverse-square approach of decoherence rates to their asymptotic values, as a function of separation distance. Inasmuch as the first-order coherence function of ideal gas in the Maxwell-Boltzmann regime has a Gaussian fall-off, the merely inverse-square saturation of decoherence rates is somewhat surprising.

\begin{figure}
\begin{center}
\includegraphics[width=0.5\textwidth]{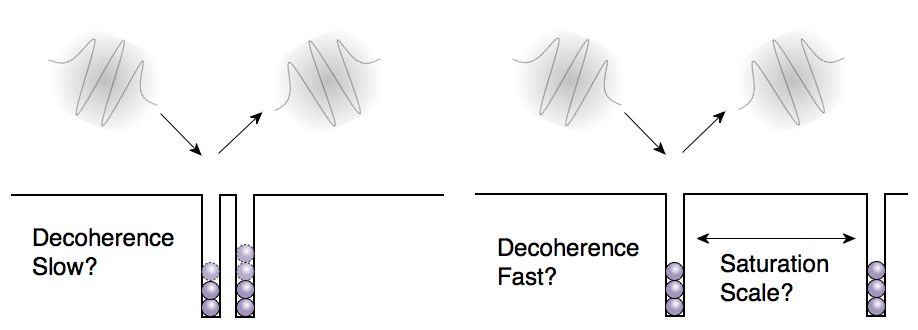}
\end{center}
\caption{Decoherence resolution length scale. Bosons of one type are coherently distributed between two tight single-mode traps. Surrounding reservoir particles interact with them by scattering, causing decoherence between different distribution states. At what separation length scale does the decoherence rate saturate?}
\label{fig:system}
\end{figure}

To check whether the somewhat counter-intuitive Born-Markov result may be a failure of the Born-Markov approximation, we then determine the asymptotic behavior at late times of the full system's exact evolution, and show that the Born-Markov result is indeed correct, in the limit of weak scattering. While the somewhat lengthy analysis of this more general calculation ultimately only confirms the conclusions about length scales that were formed from the simpler Born-Markov theory, it also supplies a rigorous treatment of collisional decoherence that may be applied beyond weak scattering.

Finally we add a finite coherence lifetime within the buffer gas reservoir, associated with a finite collisional mean free path. This provides an exponential saturation of decoherence rates with distance, but on the scale of the buffer gas mean free path scale rather than its coherence length. We conclude our paper with a brief discussion, and attach a pair of technical appendices, one on the bound states of a pair of Fermi-Huang pseudo-potentials, and one on details of the asymptotic evaluation of our exact time evolution.

\section{Model Description}\label{Model}
We consider a total system consisting of two different kinds of particles, whose populations will be regarded as two subsystems, interacting with each other. The test subsystem, in which we will study decoherence, is composed of trapped bosons, able to occupy one of two tight trapping wells at two different positions. The reservoir subsystem, which will induce the decoherence, is an ideal gas of $N_{B}$ particles within a large volume $V$, thus having density $n_{B}=N_{B}/V$; we refer to this reservoir subsystem as the `buffer gas'. The total Hamiltonian will have the basic form
\begin{equation}\label{eq:HH}
\hat{H}=\hat{H}_{S} + \hat{H}_{B} + \hat{H}_{I}\;,
\end{equation}
where the three terms govern the trapped boson system, the buffer gas reservoir, and their interaction, respectively.

We idealize the test system's two tight traps as single modes, populated in second quantization, with negligible spatial extent. The operators $\hat{n}_{\pm}$ count the number of bosons at each site $\pm$, located at positions $\vec{r}=\pm \vec{s}/2$. The separation $s=|\vec{s}|$ between the two trapping sites will be the first important length scale in our problem. We assume that our two traps are so tightly confining that no escape from them is possible, either by tunneling or by any other means, and so our test system has the Hamiltonian
\be
\hat{H}_{S}=E_{+}(\hat{n}_{+})+E_{-}(\hat{n}_{-})\label{eq:HS}
\ee
for some energy functions $E_{\pm}$. 

The very important feature of our $\hat{H}_{S}$ is that it is not important at all. The precise form of $E_{\pm}$ will have no effect whatever on the decoherence effects we compute in this paper, though their nonlinearity will generate the dispersion in the relative phase between the two bound modes that is often inaccurately called phase diffusion, and hence sometimes confused with decoherence. The point of our simple model for the system bosons has been precisely to make $\hat{H}_{S}$ so trivial that it really does nothing at all in the evolution we consider. We explain briefly why this is so, but the conclusion will be simply that we can effectively set $\hat{H}_{S}$ to zero.

\subsection{Ignoring the system Hamiltonian}
By working in the interaction picture \cite{cohen_quantum_1997} we can entirely eliminate $\hat{H}_{S}$ from the discussion, and never even mention it again. We do this in the standard textbook way, by relating the total density operator in the standard Schr\"odinger picture $\hat{\rho}_{\mathrm{Sch}}$ to the interaction picture total density operator $\hat{\rho}_{\mathrm{ttl}}$ with
\begin{equation}
\hat{\rho}_{\mathrm{ttl}}(t) = e^{\frac{i}{\hbar}\hat{H}_{S}t}\hat{\rho}_{\mathrm{Sch}}(t)e^{-\frac{i}{\hbar}\hat{H}_{S}t}
\end{equation}
The time-dependent Schr\"odinger equation for the interaction picture state operator then becomes
\begin{eqnarray}
i\hbar \frac{d}{dt}\hat{\rho}_{\mathrm{ttl}} &=& [\hat{H}_{IP}(t),\hat{\rho}_{\mathrm{ttl}}]
\end{eqnarray}
for the interaction picture Hamiltonian
\begin{eqnarray}
\hat{H}_{IP}(t) &=& e^{\frac{i}{\hbar}\hat{H}_{S}t}\left(\hat{H}-\hat{H}_{S}\right)e^{-\frac{i}{\hbar}\hat{H}_{S}t}\nonumber\\
&\equiv& \hat{H}_{B}+\hat{H}_{I}\;\;, \quad \hbox{\bf independent of $t$}
\end{eqnarray}
in our case, because both $\hat{H}_{B}$ and  $\hat{H}_{I}$ will commute with $\hat{H}_{S}$ (see the next subsection). In other words, by working in the interaction picture with any $\hat{H}_{S}$ of our chosen form, we can without loss of generality drop $\hat{H}_{S}$ completely.

The usual kind of dynamics, with energy gained or lost or transported, will not be relevant for our system bosons at all. Our only concern will be the quantum decoherence of initial states that have somehow been prepared as superpositions between different eigenstates of $\hat{n}_{\pm}$. In other words, we investigate only the quantum mechanical question of how the bath gains quantum information about the distribution of the system bosons. While our scenario is thus highly idealized, it is a generic feature of natural quantum systems that decoherence due to quantum information exchange can occur on much shorter time scales than energy transport. In this sense our analysis will be qualitatively informative about much more realistic cases: it represents a limiting case, but not just an irrelevant fiction.

\subsection{Buffer gas reservoir and interaction Hamiltonians}
For the reservoir that will cause decoherence, we assume that the buffer gas is hot and dilute enough to be described by Maxwell-Boltzmann statistics, with effectively distinguishable particles. We will therefore describe it in first quantization, with the Hamiltonian in position representation
\begin{eqnarray}\label{eq:HB}
	\hat{H}_B = \sum_{i=1}^{N_B}-\frac{\hbar^{2}}{2m}\nabla_{i}^{2}
\end{eqnarray}
where $\nabla_{i}^{2}$ is the Laplacian with respect to the position $\vec{x}_{i}$ of the $i$-th buffer gas particle, $m$ is the mass of each particle and there are $N_B$ particles in the gas.

The interaction between the buffer gas and the two bosonically populated trapping sites is due to two-particle s-wave collisions between the trapped bosons and the buffer gas particles, represented in first-quantized position representation (for the buffer gas) and second quantization (for the trapped bosons) through the Fermi-Huang pseudo-potential Hamiltonian \cite{huang_1987}
\be\label{eq:HI}
\hat{H}_{I}&=&  \sum_{\pm}\hat{n}_{\pm}\sum_{i=1}^{N} U(\vec{r}_{i}\mp \vec{s}/2)\nonumber\\
U(\vec{r}) &=& 2\pi\frac{\hbar^{2}}{m}a\,\delta(\vec{r}\,)\frac{d}{dr}r \;,
\ee
for s-wave scattering length $a$ (the next significant length scale in our problem). We are hereby assuming that the two trapped modes occupied by our system bosons are so spatially narrow compared to the wavelengths of momentum transfer in collisions with buffer gas particles that we can apply the Fermi-Huang pseudopotential for scattering, even when the scattering particles are slightly delocalized over their traps' bound modes. Because the buffer gas `sees' the strength of each of the two point-like scattering sites as proportional to the occupation number $\hat{n}_{\pm}$ of system bosons, scattering of buffer gas particles off the trapped bosons will effectively `measure' the number of bosons at each site, even though the trapped bosons are not affected by the collisions in any other way: $\hat{H}_{S}$ commutes with $\hat{H}_{I}$, just as it obviously does with $\hat{H}_{B}$, as we stated in the previous subsection.

The definition of our problem is completed by specifying that we will consider total initial states that are tensor products of bath and system states, with the bath state given by the canonical ensemble
\begin{eqnarray}\label{}	\hat{\rho}_B&=& \prod_{i=1}^{N_{B}}\hat{\rho}_{i}(\beta)\nonumber\\
\hat{\rho}_{i}(\beta)&=&Z^{-1}\sum_{\vec{k}} e^{-\beta\frac{\hbar^2k^2}{2m}}|\vec{k}\rangle\langle\vec{k}|_{i}\;.
\end{eqnarray}
Here $\beta = 1/(k_{B}T)$ is the inverse temperature of the buffer gas, $Z$ is the normalization factor, and $|\vec{k}\rangle$ are the usual plane-wave eigenstates of $\hat{H}_{B}$, with wave numbers $\vec{k}$ from the discrete set consistent with periodic boundary conditions in the large but finite volume $V$. The thermal de Broglie wavelength (or `thermal length' or `coherence length')
\begin{eqnarray}\label{}
	\lambda=\sqrt{\frac{\beta\hbar^2}{2m}}
\end{eqnarray}
is the third significant length scale in our problem. We assume that the buffer gas volume $V$ is large enough in every dimension compared to $\lambda$ that we may approximate suitably convergent sums over $\vec{k}$ as integrals, according to
\begin{eqnarray}\label{}
	\sum_{\vec{k}}\longrightarrow \frac{V}{(2\pi)^{3}}\int\!d^{3}k\;.
\end{eqnarray}
In particular this implies the normalization factor
\begin{eqnarray}\label{}
	Z \longrightarrow \frac{V}{(2\pi)^{3}}\int\!d^{3}k\,e^{-\frac{\hbar^{2}k^{2}}{2m}\beta} = \frac{V}{(4\pi)^{3/2}\lambda^{3}}\;.
\end{eqnarray}

Near the end of this paper we will consider the effects of collisions among the buffer gas particles themselves, giving rise to a fourth relevant length scale, namely the mean free path $L$. Initially, and throughout most of this paper, however, we will treat the buffer gas particles as interacting only with the trapped bosons, and not with each other.

\section{Quantum Coherence in the Buffer Gas Reservoir}

The first-order spatial coherence of a finite temperature ideal gas in the Maxwell-Boltzmann limit is Gaussian:
\begin{eqnarray}\label{eq:boltzmanncorr} g_{1}(\vec{r},\vec{r}\,') &=& \langle \vec{r}\,'|\hat{\rho}(\beta)|\vec{r}\,\rangle\nonumber\\
 &=& Z^{-1}\sum_{\vec{k}}e^{-\frac{\hbar^{2}k^{2}\beta}{2m}}e^{i\vec{k}\cdot(\vec{r}-\vec{r}\,')}\nonumber\\&{\longrightarrow}\atop{V\to\infty}& e^{-\frac{(\vec{r}-\vec{r}\,')^2}{2\lambda^2}}\;,\end{eqnarray}
where $|\vec{r}\rangle$ is an eigenstate of particle position. The characteristic length scale of this function is the thermal length $\lambda$, and it decays rapidly to zero on this scale.

Although the literal meaning of $g_{1}(\vec{r},\vec{r}\,')$ is thus quite specific, it is common to interpret it rather generally. In some sense it represents the reservoir's spatial resolution limit; the size of $g_{1}$ is a sort of measure of the gas' ignorance as to whether $\vec{r}$ and $\vec{r}\,'$ are really different points. 

Decoherence happens when an unobserved environment acquires `which way' information about the alternatives in a quantum superposition, and we can expect that scattering will express such information. In our problem of decoherence among different particle distributions between two sites separated by a distance $s=|\vec{s}|$, therefore, one might expect that the ratio $s/\lambda$ should be a decisive parameter for decoherence of the trapped boson number distribution due to buffer gas scattering.  If the separation $s$ between the two microtraps is more than a few thermal lengths, then the Gaussian fall-off of $g_{1}$ suggests that the buffer gas should be able to fully resolve the two traps as distinct scattering centers, just as if they were infinitely far apart; we would expect superposed states with the same total number of trapped particles distributed differently between the distant sites to decohere just as fast as superpositions of different total numbers in single isolated sites. When the two trapped populations are located within a thermal length of each other, however, we can expect the buffer gas environment to `see' them mainly as a single scatterer, and hence induce minor decoherence between superpositions of different number distributions between the two sites.

Based on our knowledge of $g_{1}$, we can even anticipate more precisely that the approach of decoherence rates to their asymptotic form at large separation $s\gg\lambda$ will be Gaussian, just like $g_{1}$. In fact this anticipation, reasonable as it may sound, turns out not to be accurate. Decoherence is not quite that simple. We will need to consider it as a non-trivial non-equilibrium process.

\section{Born Markov Treatment}

A Born-Markov Master equation \cite{Gardiner_Quantum_2004} for the reduced density operator $\hat{\rho}(t)$ of the system in the interaction picture is obtained by evaluating 
\be
\frac{d}{dt}\hat{\rho} = -\frac{1}{\hbar^2}\int\limits_0^t \!\!dt'\; \text{Tr} \left(\left[\hat{H}_{I'}(t),\left[\hat{H}_{I'}(t'),\hat{\rho}_{B}\otimes\hat{\rho}(t)\right]	\right]\right),\nonumber\\\label{eq:rhotime}
\ee
where the trace is over the reservoir Hilbert space. We use $\hat{H}_{I'}$ to denote a modified version of the interaction Hamiltonian, from which the expectation value of $\hat{H}_{I}$ has been subtracted out:
\begin{eqnarray}\label{}
	\hat{H}_{I'} = \hat{H}_{I}-\text{Tr}[\hat{\rho}_{B}\hat{H}_{I}]\;.
\end{eqnarray}
Note that the expectation value term that has been subtracted from $\hat{H}_{I}$ remains an operator in the system Hilbert space, since the trace is only over the reservoir sector. This term is not simply neglected, but is rather transferred into the system Hamiltonian 
\begin{eqnarray}\label{}
	\hat{H}_{S}\to \hat{H}_{S'} = \hat{H}_{S}+\text{Tr}[\hat{\rho}_{B}\hat{H}_{I}]\;,
\end{eqnarray}
so that the total $\hat{H}$ remains unchanged. The point of this procedure is just to extract that part of the reservoir's effect on the system which can be expressed as a Hamiltonian correction, leaving the purely non-Hamiltonian parts of the master equation to be derived by the Born-Markov formula above, using $\hat{H}_{I'}$. In fact, in our case, the new terms in $\hat{H}_{S'}$ can just be absorbed by redefining the energy functions $E_{\pm}$, and then by working in the slightly revised interaction picture, we again leave them with no visible effect on any of our results, and can simply ignore them.

We can express $\hat{H}_{I'}$ explicitly in the plane-wave basis that is most convenient for our buffer gas's initial state:
\be
\hat{H}_{I'}(t) &=& \sum_{i=1}^{N} \hat{H}_{i}(t)\nonumber\\
\hat{H}_{i}(t)&=& \sum_{\vec{k}\not=\vec{k}'}|\vec{k}'\rangle\langle\vec{k}|_{i}\,H_{I\vec{k}\vec{k}'}e^{-i\frac{\hbar(k^2-k'^2)}{2m}t}\nonumber\\
H_{I\vec{k}\vec{k}'}&=&\sum_{\pm}\hat{n}_{\pm}\int\!d^{3}r,\frac{e^{-i\vec{k}'\cdot\vec{r}}}{\sqrt{V}}U(\vec{r}\mp\vec{s}/2)\frac{e^{i\vec{k}\cdot\vec{r}}}{\sqrt{V}}\nonumber\\
&=&\frac{2\pi\hbar^{2}a}{mV} \sum_{\pm}\hat{n}_{\pm}  e^{\pm i\frac{(\vec{k}-\vec{k}')\cdot\vec{s}}{2}}\;.
\label{eq:HIt}
\ee
Since we have made the expectation value of $\hat{H}_{I'}$ vanish identically, and all the $\hat{H}_{i}$ commute with each other, the only non-zero contributions to the trace in (\ref{eq:rhotime}) come from the $N_{B}$ cases where two $\hat{H}_{i}(t)$ terms with the same $i$ appear together. Since all the $N_{B}$ buffer gas particles have the same mass and temperature, these $N_{B}$ non-zero contributions are all the same, and the interaction-picture $\frac{d}{dt}\hat{\rho}$ for all $N_{B}$ particles is simply $N_{B}$ times the single particle result.

The Born-Markov master equation is intended to yield the long-term effects of a weakly coupled bath, after any initial transient behavior has died away. One therefore evaluates (\ref{eq:rhotime}) in the long time limit, where the time integral yields an energy-conserving $\delta$-function just as in the textbook derivation of Fermi's Golden Rule \cite{Alicki_the_1977}, and the master equation takes the form
\be\label{eq:master}
\dot{\hat{\rho}} &=& -2\pi \frac{mN_B}{\hbar^{3}}\left(\frac{2\pi a \hbar^{2}}{mV}\right)^{2}\sum_{\vec{k}\neq\vec{k}'\atop \pm,\pm'}
	\delta\left(k^2-k'^2\right) \frac{e^{-\frac{\beta\hbar^{2}k^{2}}{2m}}}{Z}\nonumber\\
	&&\times e^{i(\pm-\pm')(\vec{k}-\vec{k}')\cdot\vec{s}/2} \left\{\left[\hat{n}_{\pm},\hat{n}_{\pm'}\hat{\rho}\right]- \left[\hat{n}_{\pm},\hat{\rho}\hat{n}_{\pm'}\right]\right\}\nonumber\\
&\to& -2\sqrt{\pi}\,n_{B} \frac{\hbar a^2}{m\lambda}\int_{0}^{\infty}\!d\xi\,\xi^{3}\,e^{-\xi^{2}}\nonumber\\
&& \qquad\times \sum_{\pm,\pm'}\left|\int_{0}^{\pi}\!\sin\theta\,d\theta\, e^{i\frac{s}{\lambda}\xi\frac{\pm-\pm'}{2}\cos\theta}\right|^{2}\nonumber\\
&&\qquad\qquad\times\left\{\left[\hat{n}_{\pm},\hat{n}_{\pm'}\hat{\rho}\right]- \left[\hat{n}_{\pm},\hat{\rho}\hat{n}_{\pm'}\right]\right\}\;.
\ee

\begin{figure}
\begin{center}
\includegraphics[width=0.45\textwidth]{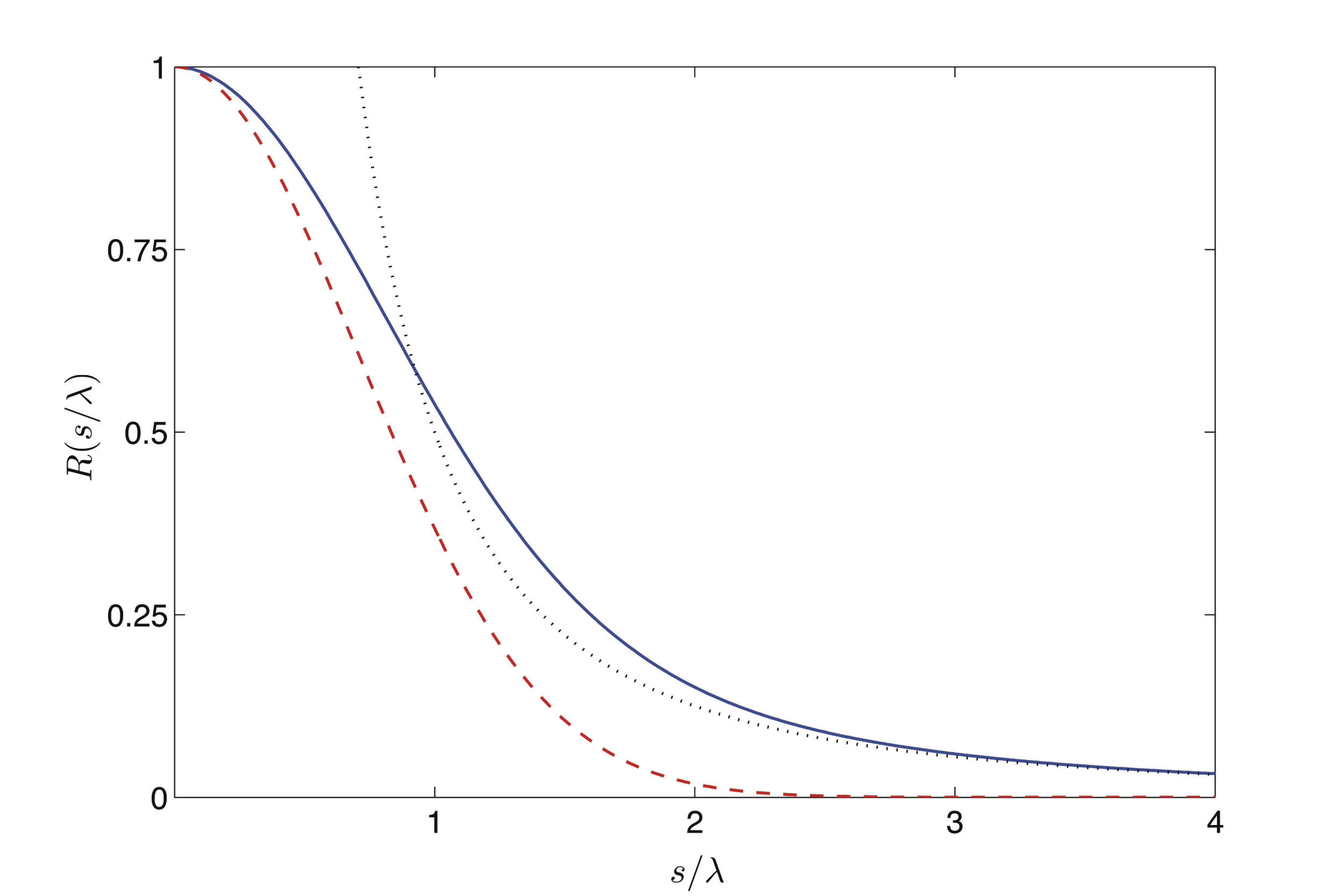}
\end{center}
\caption{Rate function $R(x)$ versus separation between the two scattering centers in units of reservoir coherence length, $x=s/\lambda$. Blue solid line: The Born-Markov decoherence rate coefficient $R(x)$. At large $x$, $R(x)$ falls off only as $1/(2x^{2})$ (black dotted line), instead of with the much more abrupt $e^{-x^{2}}$ decay (red dashed line) of spatial coherence in the thermal gas between any two points separated by a distance $x\lambda$.}
\label{fig:Rplot}
\end{figure}

We can express this master equation concisely by defining the scattering rate
\begin{eqnarray}\label{GammaDef}
\Gamma = 2\sqrt{\pi} \,n_B \frac{\hbar a^2}{m\lambda}\;.
\end{eqnarray}
Defining the total and relative particle numbers $\hat{N}=\hat{n}_++\hat{n}_-$ and $\hat{n}=\hat{n}_+-\hat{n}_-$ respectively, this becomes
\be\label{eq:me}
\dot{\hat{\rho}} &=& - \Gamma[1+R(s/\lambda)]\left(\hat{N}^2\hat{\rho}+\hat{\rho}\hat{N}^2-2\hat{N}\hat{\rho}\hat{N}\right)\nonumber\\
&& - \Gamma[1-R(s/\lambda)]\left(\hat{n}^2\hat{\rho}+\hat{\rho}\hat{n}^2-2\hat{n}\hat{\rho}\hat{n}\right)\nonumber\\
\ee
where the function $R(s/\lambda)$ is given by
\begin{eqnarray}\label{Rdef}
	R(x) &=& \frac{2}{x^{2}}\int_{0}^{\infty}\!d\xi\,\xi\,\sin^{2}(x\xi)\,e^{-\xi^{2}}\;.
\end{eqnarray}
$R(x)$ is plotted in Fig.~\ref{fig:Rplot}.  Note that $R(0)\equiv 1$ while 
\begin{eqnarray}\label{Rlimit1}
	\lim_{x\to\infty}R(x) = \frac{1}{2x^{2}}\;.
\end{eqnarray}

The master equation (\ref{eq:me}) clearly describes decoherence in the number basis for the trapped bosons. For any initial state represented in the number eigenstate basis, the exact solution of (\ref{eq:me}) is
\begin{eqnarray}\label{eq:rhosimp}
\hat{\rho}(t)&=& \sum_{N,n}\sum_{N',n'}|N,n\rangle\langle N',n'|\; \rho_{Nn,N'n'}(0)\nonumber\\
&&\times\; e^{-(1+R)\Gamma (N-N')^{2}t}\; e^{-(1-R)\Gamma (n-n')^{2}t}\;,
\end{eqnarray}
where there is no time evolution from $\hat{H}_{S}$ because we work in the interaction picture. The density matrix becomes diagonal in the number basis, but the rates at which coherence between states of different total and relative number decays are different in general, depending on $s/\lambda$.

Since $R(0)=1$, we see that the buffer gas indeed fails to resolve the two sites when their separation $s$ is much less than the gas coherence length $\lambda$. In this case $1+R\to 2$ while $1-R \to 0$, and superpositions of different total numbers of trapped bosons decohere while the subspaces of different distributions of $N$ bosons between the two sites are decoherence-free. On the other hand $R(s/\lambda)$ decays to zero for $s\gg\lambda$, and so for two micro-traps very far apart within the same thermal bath, particle number decoherence is indeed the same as if the two traps were interacting with entirely separate baths. Since $R$ depends on separation $s$ only through the ratio $s/\lambda$, it is also clear that the thermal coherence length does indeed `set the scale' for saturation of the relative number decoherence rate. 

It is somewhat surprising, however, that the approach to the separate-bath limit is not more rapid than a mere inverse square decay. Based on the first-order coherence function $g_1$, one might expect a much sharper Gaussian decay, instead.  As Fig.~\ref{fig:Rplot} shows, the actual decay of $R(x)$ is significantly though perhaps not drastically slower than Gaussian. The decay of $R(x)$ is probably not sufficiently gentler than expected to be important for things like providing decoherence-free subspaces for quantum computing, but it should be observable from interference fringe contrasts or shot-to-shot variation in fringe positioning. And the non-Gaussian nature of $R(x)$ does tell us something basic and important about decoherence. 

The first-order coherence function $g_1(\vec{r},\vec{r}\,')$ is a property of the unperturbed thermal state. Decoherence of a scattering target under bombardment by gas particles, however, is a phenomenon of non-equilibrium time evolution. It involves not only integration over the initial state of the reservoir, in its thermal ensemble, but also the trace over the reservoir's final state. The initial and final states of the reservoir are of course related by unitary time evolution, so these two integrations over states are not independent. 

Following through our master equation's derivation, it is possible to see that the terms which acquire $R(s/\lambda)$ factors come from phases $e^{\pm i(\vec{k}-\vec{k}')\cdot\vec{s}}$ that are averaged over both $\vec{k}$ and $\vec{k}'$. If these two integrations were fully independent, their results would both be very small for large $s/\lambda$, as the oscillating integrands averaged away. In fact, however, time evolution conserves kinetic energy, and so $|\vec{k}'|=|\vec{k}|$. 
This means that, after integrating independently over the two solid angles in $\vec{k}$- and $\vec{k}'$-space, instead of two independent $k$- and $k'$-averages of rapidly oscillating sinusoidal factors $\propto \sin(ks)\sin(k's)$, we have a single thermal average over a factor $\propto2\sin^{2}(ks) = 1-\cos(2ks)$, whose first term does not oscillate and hence does not wash out for large site separation $s$.

If we think of the thermal ensemble as a random choice of wave number for each particle, then what this says is that the instantaneous relative phase between points $\pm\vec{s}/2$ becomes nearly random for each particle, once $s\gg\lambda$, so the ensemble average for $g_1$ approaches zero; but the phase that appears in the evolution of each particle between initial and final states, with scattering off two different sites at $\pm\vec{s}/2$, has a systematic component whose ensemble average in $R$ is not nearly so small. 

Another way to look at decoherence of the two sites by scattering is to switch notation from $N,n$ to $n_\pm$, and re-write (\ref{eq:rhosimp}) as
\begin{eqnarray}\label{eq:rhotentang}
\hat{\rho}(t)&=& \sum_{n_{+}n_{-}}\sum_{n'_{+},n'_{-}}|n_+, n_-\rangle\langle  n'_+, n'_-|\; \rho_{n_+ n_-, n'_+ n'_-}(0)\nonumber\\
&&\times\; e^{-2\Gamma [(n_{+}-n'_{+})^{2}+(n_{-}-n'_{-})^{2}]t}\nonumber\\
&&\times\;e^{-4R\Gamma (n_{+}-n'_{+})(n_{-}-n'_{-})t}\;.
\end{eqnarray}
This shows that the buffer gas reservoir effectively `measures' the number of bosons in each site $\pm$, at a rate which is independent of the separation $s$ between the sites; but it also induces entanglement between the two sites. The exponential containing $R$ actually tends to increase the probability of boson number fluctuations when they are anti-correlated between the sites (though it is always outweighed by the other term, and manages only to slow down the rate at which probability for these fluctuations is lost). 

It is interesting to see how the Born-Markov formalism, which is most often used to describe more familiar forms of damping and noise and decoherence, is in fact quite able to capture this entanglement-building aspect of the reservoir's evolution. In the second-order perturbative description used for the Born-Markov equation, the inter-site entanglement arises from two sources: quantum interference between two first-order processes in which a reservoir particle scatters from two different sites; and interference between a zeroth order process (no scattering) and a second-order process in which a particle scatters from one site, and then from the other, and ends up with exactly its initial momentum. 

The important and general point here is indeed that decoherence is not a phenomenon of instantaneous equilibrium, but of non-equilibrium time evolution. The fact that one cannot accurately predict decoherence entirely from such equilibrium properties as $g_1(\vec{r},\vec{r}\,')$ can be considered a quantum counterpart to Landauer's warning that many different kinds of evolution can produce the same equilibrium, so that non-equilibrium evolution can never be inferred from the equilibrium state \cite{landauer_motion_1988}.

Our expectation in Section III that decoherence should saturate with a Gaussian dependence on separation, because $g_1$ is Gaussian, was really nothing but an attempt to make bricks without straw, by deducing the non-equilibrium phenomenon of decoherence from an equilibrium property. The Born-Markov master equation, however, is in a sense only a little bit better than that; it tries to make bricks with only a little bit of straw. It uses second-order time-dependent perturbation theory to describe non-equilibrium evolution, and hence assumes that the reservoir state remains close to equilibrium at all times. For the simple model that we have taken in this paper, however, we can take all the straw we need to make good bricks: we can solve the reservoir's full time evolution exactly. For the asymptotic decoherence rate at late times, which is all that matters for a dilute gas reservoir whose effects are only significant over long times, we can provide an explicit formula.\\

\section{Exact solution for an ideal buffer gas}
\subsection{Reduction to a single-particle problem}\label{BathOneParticle}
The time evolution of the total system-bath density matrix, from the initial tensor product state, has the formal exact solution in the interaction picture
\be
\hat{\rho}_{\mathrm{ttl}}(t)= e^{-\frac{i}{\hbar}\hat{H}_{IP}t}[\hat{\rho}(0)\otimes\hat\rho_{B}]e^{\frac{i}{\hbar}\hat{H}_{IP} t}\;,\label{eq:solformal}
\ee
with $\hat{H}_{IP} = \hat{H}_{B}+\hat{H}_{I}$ as given by (\ref{eq:HB}) and (\ref{eq:HI}). \textbf{We will now drop the subscript} $_{IP}$ and simply write $\hat{H} = \hat{H}_{B}+\hat{H}_{I}$, as if we had merely set $\hat{H}_{S}$ to zero from the start.\\

The reduced density matrix of the trapped boson system is then given in the basis of eigenstates of $\hat{n}_{\pm}$ as follows (we will explain each line below):
\begin{widetext}
\begin{eqnarray}
\rho_{\vec n \vec n'}(t) &=& \sum_{\vec m \vec m'}\rho_{\vec m \vec m'}(0)\mathrm{Tr}\left[\langle \vec n|e^{-\frac{i}{\hbar}\hat{H}t}\left(|\vec m\rangle\langle \vec m'|\otimes \hat{\rho}_{B}\right)e^{\frac{i}{\hbar}\hat{H}t}|\vec n'\rangle\right]
\nonumber\\
&\equiv& \rho_{\vec n \vec n'}(0)\mathrm{Tr}\left[\langle \vec n|e^{-\frac{i}{\hbar}\hat{H}t}|\vec n\rangle\,\hat{\rho}_{B}\,\langle \vec n'|e^{\frac{i}{\hbar}\hat{H}t}|\vec n'\rangle\right]\nonumber\\
&\equiv& \rho_{\vec n \vec n'}(0)\mathrm{Tr}\left[e^{-\frac{i}{\hbar}\hat{H}_{\vec n}t}\,\hat{\rho}_{B}\,e^{\frac{i}{\hbar}\hat{H}_{\vec n'}t}\right]\nonumber\\
\hat{H}_{\vec n}t &\equiv& \sum_{i=1}^{N_{B}}\left[-\frac{\hbar^{2}}{2m}\nabla_{i}^{2} + \sum_{\pm} n_\pm U(\vec{r}_{i}\mp \vec{s}/2)\right]\;.
\end{eqnarray}
\end{widetext}
Here we use $\vec n$ as a short form for the doublet $(n_{+},n_{-})$. In the first line we have simply taken the reduced density operator of the boson system by partially tracing $\hat{\rho}_{\mathrm{ttl}}$ over the buffer gas reservoir sector of the total Hilbert space, and expressed that reduced density operator in the basis of $\hat{n}_{\pm}$ eigenstates, at both the final time $t$ and the initial time $0$. In the second line we have noted, from the fact that $\hat{H}$ commutes with $\hat{n}_{\pm}$, that only the case $\vec m=\vec n$, $\vec m'=\vec n'$ will not vanish. Then in the third and fourth lines we have simply seen that sandwiching the $e^{\pm\frac{i}{\hbar}\hat{H}t}$ operators between eigenstates of $\hat{n}_{\pm}$ is equivalent to replacing the operator $\hat{n}_{\pm}$ in $\hat{H}$ with the corresponding eigenvalue $n_\pm$. All of the steps above are exact.

Since $\hat{\rho}_{B}$ is a direct product over the $N_{B}$ identical buffer gas particles, and $\hat{H}_{\vec n}$ is a sum over them, we have reduced our evolution problem to that of only a single particle, in the following sense:
\begin{eqnarray}\label{eq:F1}
\rho_{\vec n \vec n'}(t) &\equiv& \rho_{\vec n \vec n'}(0) \times \left(F(\vec n,\vec n',t)\right)^{N_{B}}\nonumber\\
F(\vec n,\vec n',t) &\equiv&\sum_{\vec{k}} \frac{e^{-\frac{\beta \hbar^{2}k^{2}}{2m}}}{Z} \int\!d^{3}r\,  \Psi^{*}_{\vec{k}}(\vec{r},\vec n',t)\Psi_{\vec{k}}(\vec{r},\vec n,t)\;,\nonumber\\
\end{eqnarray}
where $\Psi_{\vec{k}}(\vec{r},\vec n,t)$ is defined by the initial condition
\begin{equation}\label{eq:scatinitialcondition}
\Psi_{\vec{k}}(\vec{r},\vec n,0) =  \Psi_{\vec{k}}(\vec{r})=\frac{e^{i\vec{k}\cdot\vec{r}}}{\sqrt{V}}
\end{equation}
for $\vec{k}$ satisfying periodic boundary conditions in the volume $V$ and the single-particle Schr\"odinger evolution
\begin{equation}\label{1PSE}
i\hbar\frac{\partial}{\partial t}\Psi_{\vec{k}}(\vec{r},\vec n,t) =  \left[-\frac{\hbar^{2}}{2m}\nabla^{2} + \sum_{\pm} n_\pm U(\vec{r}_{\pm})\right]\Psi_{\vec{k}}(\vec{r},\vec n,t)\;,
\end{equation}
where we write $\vec{r}_{\pm}=\vec{r}\mp \vec{s}/2$ for brevity.\\

What we see is that the decoherence of superpositions of different $\vec n$ distributions is given exactly in our model in terms of the inner product between two states that evolve, from the same initial state, under different Hamiltonians. To the extent that the different trapped boson populations drive the buffer gas into orthogonal states, the different boson states will decohere. For a buffer gas reservoir composed of identical non-interacting particles, the decoherence factor simply scales exponentially with the number of reservoir particles. The only non-trivial element to be understood is a single-particle problem, of scattering from two pointlike sites separated by the distance $s$.

\subsection{Exact single-particle eigenstates}\label{Eigenstates}

Our approach will be to solve the single-particle time evolution problem (\ref{1PSE}) within an infinite volume, without periodic boundary conditions, for any value of the initial wave number $\vec{k}$. We will then specialize to the case of periodic solutions within the finite volume $V$, by restricting the initial $\vec{k}$ to appropriate discrete values, and ensuring that our later time evolution does not violate the periodicity. Within the infinite volume, therefore, Eqn.~(\ref{1PSE}) can as usual be solved as an integral over eigensolutions 
\begin{eqnarray}\label{eq:eigensol}
	\Psi_{\vec{k}}(\vec{r},\vec n,t) &=& \int\!d^{3}k'\,\phi_{\vec{k}',\vec{n}}(\vec{r}\,)e^{-i\frac{\hbar k'^{2}}{2m}t}\nonumber\\
&&\times\,\int\!d^{3}r'\,\phi_{\vec{k}',\vec{n}}^{*}(\vec{r}\,')\Psi_{\vec{k}}(\vec{r}\,)
\end{eqnarray}
where the $\phi_{\vec{k},\vec{n}}$ are normalized solutions to the time-independent Schr\"odinger equation,
\be\label{TISE1}
\frac{\hbar^{2}k^{2}}{2m}\phi_{\vec{k},\vec{n}} &=& \left[-\frac{\hbar^{2}}{2m}\nabla^{2} + \sum_{\pm} 2\pi\frac{\hbar^{2}}{m}a_{\pm}\,\delta(\vec{r}_{\pm})\frac{d}{dr_{\pm}}r_{\pm}\right]\phi_{\vec{k},\vec{n}} \;.\nonumber\\
\ee
Here we write $r_{\pm}\equiv |\vec{r}_{\pm}|$ and introduce $a_{\pm} \equiv a n_{\pm}$ for both brevity and generality, since our solutions will be valid even when $a_{\pm}$ are not whole number multiples of each other.

We obtain the exact solution of (\ref{TISE1}) for an infinite volume without periodic boundary conditions by motivating a simple Ansatz, and then showing that it works. The motivation is to note that pseudo-potential scattering centers generate outgoing s-waves proportional to every incoming wave which strikes them. In principle one could therefore imagine an infinite Born series of outgoing s-waves being emitted from our two scatterers, with each pair of s-waves from one order in Born expansion generating more s-waves in the next order, as the s-wave from one scatterer encounters the other at a distance $s$. What this means, however, is precisely that the exact solution must be a three-term superposition of the form
\be\label{ScatteringAnsatz}
\phi_{\vec{k},\vec{n}} &=& \frac{1}{(2\pi)^{3/2}}\left[e^{i\vec{k}\cdot\vec{r}}-\phi^{\mathrm{scat}}\right]\;,\\
\phi^{scat}&=&\sum_{\pm}\alpha_{\pm}\frac{e^{ikr_{\pm}}}{r_{\pm}}
\ee
That is, there will be an unscattered wave plus two outgoing s-waves, centered respectively on each scattering center. The long Born series of multiple scatterings would simply be a way of computing the co-efficients $\alpha_{\pm}$ self-consistently. 

We can see that if the Ansatz wave functions are indeed eigenfunctions, then they are all correctly normalized as written above, for any $\alpha_{\pm}$. A proof of this normalization can be based on the fact that the Schr\"odinger inner product is invariant under time evolution. We can consider any two Gaussian superpositions of these $\phi_{\vec{k},\vec{n}}$, with different mean $\vec{k}$, and in the limit where the Gaussian weights are very narrow in $\vec{k}$-space, the inner product between the two Gaussian packets will just be the inner product between the two $\phi_{\vec{k},\vec{n}}$. But if we time evolve such Gaussian packets back to $t\to-\infty$, the superposition integral over $\vec{k}$ can be performed by the method of steepest descents (because $|t|$ is so large), and there will be no contribution from the outgoing s-wave terms. This shows that the first, unscattered wave term in (\ref{ScatteringAnsatz}) is the only one relevant for normalization, and the $(2\pi)^{-3/2}$ prefactor is therefore correct. 

To determine the correct $\alpha_{\pm}$ coefficients, instead of computing the Born series, we can simply insert the Ansatz (\ref{ScatteringAnsatz}) into the time-independent Schr\"odinger equation (\ref{TISE1}) and solve for $\alpha_{\pm}$. For this we use the basic results for the pseudopotential \cite{huang_1987}

\be
[\nabla^{2}+k^{2}]\frac{e^{ikr_{\pm}}}{{r}_{\pm}}&=& -4\pi\delta(\vec{r}_{\pm})\nonumber\\
U(\vec{r}_{\pm})\, e^{i\vec{k}\cdot\vec{r}} &=& \frac{2\pi \hbar^2 a_\pm}{m}\,\delta(\vec{r}_{\pm})\,e^{i\vec{k}\cdot\frac{\vec{s}}{2}}\nonumber\\
U(\vec{r}_{\pm})\, \frac{e^{ik{r}_{\mp}}}{{r}_{\mp}}&=&\frac{2\pi \hbar^2 a_\pm}{m}\,\delta(\vec{r}_{\pm})\,\frac{e^{iks}}{s}\nonumber\\
U(\vec{r}_{\pm})\, \frac{e^{ik{r}_{\pm}}}{{r}_{\pm}}&=&\frac{2\pi \hbar^2 a_\pm}{m}\,\delta(\vec{r}_{\pm})\,ik\;.\nonumber
\ee
These results imply that in order for (\ref{ScatteringAnsatz}) to solve (\ref{TISE1}), the total coefficients of both delta functions $\delta(\vec{r}\mp \vec{s}/2)$ must separately vanish. In particular, the $\alpha_{\pm}$ must together satisfy the coupled linear equations
\be\label{SchroedingerSC}
A_{\pm}\alpha_{\pm}+B_{\pm}\alpha_{\mp} &=& a_{\pm}\,e^{\pm i\vec{k}\cdot\vec{s}/2}
\ee
for
\be\label{ABcoeff}
A_{\pm} &=& 1+ia_{\pm} k\;,\;\;\;\;\;\;
B_{\pm} = \frac{a_{\pm}}{s}e^{iks}\;.
\ee
A bit of manipulation then yields the explicit solution for the $\alpha_{\pm}$, and inserting them into the Ansatz (\ref{ScatteringAnsatz}) reveals
\be\label{eq:psieigen}
\phi_{\vec{k},\vec{n}}(\vec{r}\,) &=& \frac{1}{(2\pi)^{3/2}}\left[e^{i\vec{k}\cdot\vec{r}}-\sum_{\pm}\alpha_{\pm}(\vec{k},\vec n)\frac{e^{ik{r}_{\pm}}}{{r}_{\pm}}\right]\nonumber\\\
\alpha_{\pm}(\vec{k},\vec n) &\equiv&
	a_{\pm}\frac{A_{\mp} e^{\pm i \vec{k}\cdot\vec{s}/2}\!-\!B_{\mp}e^{\mp i \vec{k}\cdot\vec{s}/2}}{A_+A_--B_+B_-}\;.
\ee
Please note that the coefficients $A_{\pm}$ and $B_{\pm}$ are dependent on the site occupation numbers $\vec{n}$ via $a_{\pm}$ and the wave vector's absolute value $k=|\vec{k}|$.  Though for sake of notational brevity we omit spelling out this dependencies in the further discussion.

In effect (\ref{eq:psieigen}) is the central result of our paper, and the rest of our analysis will simply exploit it. Before we begin bringing this result back into our many-body decoherence problem, we can point out some of its features as a purely single-particle result. Firstly it is easy to see that in the limit $a_{\pm}/s\to0$ where the two scatterers are very far apart, $B_{\pm}\to0$ and we recover as the full, self-consistent $\phi^{\mathrm{scat}}$ simply the sum of the two scattered s-waves that would be generated by each scattering center alone:
\be
\lim_{\frac{a_{\pm}}{s}\to 0}\phi_{\vec{k},\vec{n}}^{\mathrm{scat}} = \sum_{\pm}\frac{a_{\pm}}{1+ia_{\pm}k}e^{\pm i \vec{k}\cdot\vec{s}/2}\frac{e^{ik{r}_{\pm}}}{{r}_{\pm}}\;.
\ee
In the opposite limit, however, where the two scatterers lie well within the scattering lengths of each other, $B_{\pm}$ become large, and so the scattered wave becomes small, even when the scattering lengths separately are not. In effect, two strong s-wave scatterers very close to each other somehow cancel one another. This is perhaps surprising, but it would seem to be a real effect of coherent scattering from two point-like centers. Of course, using pseudo-potentials for the two scatterers implies that the actual short-ranged potentials that cause the scattering remain well separated from each other. The model therefore fails to represent physical scattering in the strict limit $s\to 0$. Since scattering resonances can produce scattering lengths $a_{\pm}$ much longer than the range of the scattering potential, however, the limit $s\ll |a_{\pm}|$ is certainly possible in principle even with $s$ larger than the actual range of inter-particle potentials, so that our pseudo-potential model remains valid. The sharp reduction of scattering from two targets in that limit, in comparison with scattering from either target in isolation, is a curious wave-mechanical effect, whose possible implications for many-body dynamics may be interesting to consider.

It is important to note that our solutions do \textit{not} become invalid at radii from the scatterer less than $|a_{\pm}|$: by construction the Fermi-Huang pseudo-potential provides exact scattered s-waves everywhere outside the range of whatever short-ranged physical potential is providing the scattering, and in cases of scattering resonance it is perfectly possible for the scattering length to be much greater than this short range.


\subsection{Asymptotic solution at late times.}
Schr\"odinger evolutions as in (\ref{eq:eigensol}) can be approximated for large values of $t$ by performing angular integrals in $\vec{k}$-space exactly, and then evaluating the radial integral over $k$ by the method of stationary phase \cite{bender_advanced_1999}. The result is a leading order term, contributed by a saddlepoint of the integrand somewhere in the complex plane of analytically continued $k$, followed by a series of corrections at higher orders in $1/t$. Here we take the view that only the leading term is really of interest, because only it will lead to decoherence effects on the trapped boson system that grow steadily in time. The higher order terms in $1/t$ amount only to transient effects, which are in any case apt to be artifacts of the factorized initial state of system and reservoir, which was assumed for convenience rather than realism. Transient phenomena in this model are sensitive to specific details which the model is not really intended to describe. Moreover, if the buffer gas density is high enough that the transient effects are not quantitatively negligible, then the whole picture of it as a passive unobserved environment is called into question, as the environment will have a very substantial effect on the system. We wish instead to consider a situation where the environment's effect is instantaneously very weak, so that it only becomes important at all by steady accumulation over a long time. In this sense it is really the asymptotic evolution at late times that we wish to compute, and it would be inconsistent in a sense \textit{not} to neglect transients, because our model itself is not to be taken seriously for describing realistic transient phenomena.

We will therefore evaluate (\ref{eq:eigensol}) using stationary phase, and discard higher order corrections in $1/t$, as described in Appendix A. The leading order result is
\begin{eqnarray}\label{eq:statphase1}
	\Psi_{\vec{k}}(\vec{r},\vec n,t) &\doteq& \frac{1}{\sqrt{V}}\left[e^{i\vec{k}\cdot\vec{r}}-\psi^{\mathrm{scat}}(\vec{r},t)\right]\,e^{-i\frac{\hbar k^{2}}{2m}t}\\
\psi^{\mathrm{scat}}(\vec{r},t)&=&\sum_{\pm}\alpha_{\pm}\,\Theta\left(\frac{\hbar k}{m}t-{r}_{\pm}\right)\,\frac{e^{ik{r}_{\pm}}}{{r}_{\pm}}\nonumber
\end{eqnarray}
where we have dropped transient terms, that will contribute only quantities higher order by at least a factor of $1/t$ than the results we will obtain with the terms we have displayed. As we explain in Appendix B, the result (\ref{eq:statphase1}) is valid for all values of $a_{\pm}$, whether bound states exist or not. The step function $\Theta(x) $ vanishes for $x<0$ and equals one for $x>0$, so $\psi_{\mathrm{scat}}$ includes two s-waves extending from origins at $\pm\vec{s}/2$ to radius $(\hbar k/m)t$. Non-relativistic Schr\"odinger evolution does not exactly have such sharp causal wavefronts, but to leading order in $1/t$ it does: the effect of non-relativistic dispersion is just a transient broadening of the wavefront over a thickness of order $\sqrt{t}$, which ultimately contributes to all the integrals we will compute only post-leading terms in $1/t$.

If we wish we can drop further transients from our asymptotic solution by noting that for large $(\hbar k/m)t\gg s$, most of the support of our $\psi_{\mathrm{scat}}$ terms will be over $r$ so large that the difference between $r$ and $|\vec{r}\mp\vec{s}/2|$ is only a small correction:
\begin{eqnarray}\label{}
	{r}_{\pm}=|\vec{r}\mp\vec{s}/2| = r \mp \hat{r}\cdot\vec{s}/2 +\mathcal{O}(r^{-1})\;,
\end{eqnarray}
where by $\hat{r}$ we denote not an operator but the unit vector in the direction of $\vec{r}$.
We will therefore ultimately incur only errors of higher order in $1/t$ by replacing
\begin{eqnarray}\label{}
	\frac{e^{ik{r}_{\pm}}}{{r}_{\pm}}\longrightarrow \frac{e^{ikr}}{r}e^{\mp i\frac{\hat{r}\cdot\vec{s}}{2}}\;.
\end{eqnarray}

This leaves a second equally accurate asymptotic late time solution as an alternative to (\ref{eq:statphase1})
\begin{eqnarray}\label{eq:statphase2}
	\Psi_{\vec{k}}(\vec{r},\vec n,t) &\doteq& \frac{e^{-i\frac{\hbar k^{2}}{2m}t}}{\sqrt{V}}\left[e^{i\vec{k}\cdot\vec{r}}+\Theta\left(\frac{\hbar k}{m}t-r\right)\,\frac{e^{ikr}}{r}\, f_{\vec{k}}(\hat{r},\vec n)\right]\nonumber\\
	f_{\vec{k}}(\hat{r},\vec n) &=& -\sum_{\pm}\alpha_{\pm}(\vec{k},\vec n)e^{\mp i k\hat{r}\cdot\vec{s}/2}\;,
\end{eqnarray}
where $f_{\vec{k}}(\hat{r},\vec n)$ is (according to the standard textbook definition \cite{taylor_scattering_2006}) the scattering amplitude for our pair of scattering sites regarded as a single non-spherical scattering potential. The appearance of the scattering amplitude $f$ here in our time-dependent scattering problem shows that our asymptotic late-time analysis by the stationary phase approximation has in effect supplied the step which is too often glossed over in basic texts, of relating the time-independent eigenfunctions to the physical phenomenon of time-dependent scattering.
 
As long as we do not consider $t$ so large that $r=(\hbar k/m)t$ extends beyond the limits of our large volume $V$, then both of our asymptotic solutions (\ref{eq:statphase1},\ref{eq:statphase2}) still satisfy the necessary periodic boundary conditions in the finite volume, for the appropriate discrete values of $\vec{k}$. We have therefore solved (\ref{1PSE}) for our finite volume asymptotically at late times, to leading order in $1/t$; but in contrast to our Born-Markov analysis of Section IV, our solution here is non-perturbative in the interaction Hamiltonian $\hat{H}_{I}$.

\begin{widetext}
\subsection{Exact asymptotic decoherence}
Using our late time solution we can now compute the inner product in (\ref{eq:F1}) to leading order in $1/t$:
\begin{eqnarray}\label{}
	\int\!d^{3}r\,  \Psi^{*}_{\vec{k}}(\vec{r},\vec n',t)\Psi_{\vec{k}}(\vec{r},\vec n,t) &\doteq& 1 - \frac{t}{N_B}A_{\vec{k}}(\vec n,t) - \frac{t}{N_B}A^{*}_{\vec{k}}(\vec n',t) + 2\frac{t}{N_B}B_{\vec{k}}(\vec n,\vec n',t)\\
A_{\vec{k}}(\vec n)&\equiv&\frac{1}{t}\frac{N_B}{V}\sum_{\pm}\alpha_{\pm}(\vec{k},\vec n)\oint\!d^{2}\hat{r}\int_{0}^{\frac{\hbar k}{m}t}\!r^{2}dr\,e^{-i\vec{k}\cdot(\vec{r}\pm\vec{s}/2)} \frac{e^{ikr}}{r}\nonumber\\
&\equiv&-2\pi i\,n_B\frac{\hbar}{m}f(\hat{k},\vec n)\;\times [1+\mathcal{O}(1/t)] \\
B_{\vec{k}}(\vec n,\vec n')
&\equiv&\frac{n_{B}}{2}\frac{\hbar}{m}k\oint\!d^{2}\hat{r} f_{\vec{k}}(\hat{r},\vec n)f^{*}_{\vec{k}}(\hat{r},\vec n')\nonumber\\
&=&2\pi \,n_B\frac{\hbar}{m} k\sum_{\pm}\alpha_{\pm}(\vec{k},\vec n)\left[\alpha^{*}_{\pm}(\vec{k},\vec n') + \frac{\sin(ks)}{ks}\alpha^{*}_{\mp}(\vec{k},\vec n') \right]
\end{eqnarray}
Here we have introduced the factors of $t/N_B$ in the definitions of $A_{\vec{k}}$ and $B_{\vec{k}}$ so that both quantities end up as finite rates in the limits $t\to\infty$ and $N_B, V\to\infty$ for constant mean buffer gas density $n_B=\frac{N_B}{V}$. We have used the first form (\ref{eq:statphase1}) of our asymptotic solution for $\Psi_{\vec{k}}$ to evaluate $A_{\vec{k}}$, with a shift $\vec{r}\to \vec{r}\pm\vec{s}/2$ in the integration variables, and we have used the second form (\ref{eq:statphase1}) to evaluate $B_{\vec{k}}$. For consistency in keeping only leading terms in $1/t$, we need to discard the $+\mathcal{O}(1/t)$ corrections noted in our result for $A_{\vec{k}}$.

Inserting the above result into (\ref{eq:F1}), taking the limit $V\to\infty$ for fixed $n_B$, and assuming negligible change in $\hat{\rho}$ from the transient evolution that we have not expressed here, we find that at late times
\begin{eqnarray}\label{eq:exevol}
	\rho_{\vec n \vec n'}(t) &\doteq& \rho_{\vec n\vec n'}(0) \times \left[1+\frac{t}{N_B}\left[-i\Omega(\vec n)+i\Omega(\vec n') - D(\vec n,\vec n')\right]\right]^{N_B}\nonumber\\
&\longrightarrow\atop{N_B\to\infty}& \rho_{\vec n \vec n'}(0) \times e^{-it\Omega(\vec n)}e^{it\Omega(\vec n')} e^{-tD(\vec n,\vec n')}
\end{eqnarray}
for
\begin{eqnarray}\label{}
	\Omega &\equiv&\frac{\lambda^3}{\pi^{3/2}}\,\int\!d^{3}k\,e^{-\lambda^{2}k^{2}}\mathrm{Im}(A_{\vec{k}})\nonumber\\
D &\equiv& \frac{\lambda^3}{\pi^{3/2}} \,\int\!d^{3}k\,e^{-\lambda^{2}k^{2}}\left(\mathrm{Re}[A_{\vec{k}}(\vec n) + A^{*}_{\vec{k}}(\vec n')]-2B_{\vec{k}}(\vec n,\vec n')\right)\;.
\end{eqnarray}
\end{widetext}

The effect of $\Omega(\vec n)$ on $\rho_{\vec n \vec n'}(t)$ is merely that of a correction to the $E_{\pm}(\hat{n}_{\pm})$ functions in the system Hamiltonian $\hat{H}_{S}$.  Since our interest is in decoherence rather than such quite trivial Hamiltonian evolution, we will restrict our further analysis to the non-Hamiltonian term $D(\vec n,\vec n')$.

Considering our pair of scattering sites simply as a general scattering potential, the optical theorem supplies a useful identity which can also be verified directly with somewhat lengthy but straightforward algebra:
\begin{eqnarray}\label{eq:optical}
\mathrm{Re}[A_{\vec{k}}(\vec n)] &\equiv& 2\pi\, n_B\frac{\hbar}{m}\mathrm{Im}[f_{\vec{k}}(\hat{k},\vec n)]\nonumber\\
&=& 2\pi \,n_B\frac{\hbar}{m}\frac{k}{4\pi} \oint\!d^{2}\hat{r}\,|f_{\vec{k}}(\hat{r},\vec n)|^{2}\nonumber\\
&=& B_{\vec{k}}(\vec n,\vec n)\;.
\end{eqnarray}
This result allows us to express our decoherence rate $D$ as
\begin{eqnarray}\label{eq:DD}
D &=& \gamma(\vec n,\vec n)+\gamma(\vec n',\vec n')-2\gamma(\vec n,\vec n')\nonumber\\
\gamma(\vec n,\vec n')&\equiv&\pi^{-3/2} \int\!d^{3}z\,e^{-z^{2}}\, B_{\vec{z}/\lambda}(\vec n,\vec n')\;,
\end{eqnarray}
where we have introduced $\vec{z} = \lambda\vec{k}$ as a dimensionless integration variable.

By suitably grouping terms in the sums over $\pm,\pm'$ in our final expression for $B_{\vec{k}}(\vec n,\vec n')$, we can express $B_{\vec{k}}(\vec n,\vec n')$ in the form $B_{\vec{k}}(\vec n,\vec n') = G_{1\vec{k}}(\vec n)G^{*}_{1\vec{k}}(\vec n')+G_{2\vec{k}}(\vec n)G^{*}_{2\vec{k}}(\vec n')$, for certain functions $G_{j\vec{k}}$; this shows that the evolution of our reduced density matrix (\ref{eq:exevol}) obeys a master equation of Lindblad form.

Collecting and inserting all of the intermediate quantities that we have defined, we can finally express the essential decoherence rate functions explicitly in terms of our model's basic quantities:\\
\begin{widetext}
\begin{eqnarray}\label{eq:gamma}
\gamma(\vec n,\vec n') &=& 4\Gamma \int_0^\infty dz\,\frac{e^{-z^{2}}\, z^3}{\left[(1+i\frac{a}{\lambda}zn_+)(1+i\frac{a}{\lambda}zn_-)- \left(\frac{a}{s}\right)^2n_+n_-e^{2i\frac{zs}{\lambda}}\right] \left[(1-i\frac{a}{\lambda}zn'_+)(1-i\frac{a}{\lambda}zn'_-)- \left(\frac{a}{s}\right)^2n'_+n'_-e^{-2i\frac{zs}{\lambda}}\right]}\nonumber\\
&\times& \left\{2\left[ \left(\frac{a^2}{s^2}+\frac{a^2}{\lambda^2}z^2 \right)\left(1+\mbox{sinc}^2\left(\frac{zs}{\lambda}\right)\right) - 4 \,\frac{a^2}{s^2}\sin^2\left(\frac{zs}{\lambda}\right) \right] n_+n_-n'_+n'_- \right. \nonumber\\
& &+\left. i\,\frac{a}{\lambda}\,z \left(1+\mbox{sinc}^2\left(\frac{zs}{\lambda}\right)\right) \left( n_+n_-N'-N\,n'_+n'_- \right) - 2\, \frac{a}{s}\cos\left(\frac{zs}{\lambda}\right)\mbox{sinc}\left(\frac{zs}{\lambda}\right)\left( n_+n_-N'+N\,n'_+n'_- \right)\right. \nonumber\\
& & \left. + n_+n'_++n_-n'_-+\mbox{sinc}\left(\frac{zs}{\lambda}\right)\left( n_+n'_-+n_-n'_+ \right)\right\}
\end{eqnarray}
\end{widetext}
with $N=n_++n_-, \, N'=n'_++n'_-$, and $\mbox{sinc}(x)=\sin (x)/x$. Here $\Gamma = 2\sqrt{\pi} \,n_B \frac{\hbar a^2}{m\lambda}$ is the same rate we defined in (\ref{GammaDef}), for our Born-Markov calculation. Insofar as our question is about the decoherence rate in our model after negligible transients have died away, this formidable expression is the exact answer to our question. 

Unlike the Born-Markov result, Eqn.~(\ref{eq:gamma}) is not limited to the weak interaction limit, but remains valid even in the so-called unitary limit where $an_{\pm}$ becomes non-perturbatively large (whether because a scattering resonance between the trapped and buffer gas particles makes the scattering length $a$ long, or simply because the number of tightly trapped bosons $n_{\pm}$ becomes large). Results in such strong interaction limits will be considered elsewhere; in the present paper, we restrict our attention to the weak scattering limit, and regard (\ref{eq:gamma}) as a vindication of the Born-Markov approximation for this form of decoherence. For in the limit where $a/\lambda$ and $a/s$ can both be neglected entirely (weak interaction), we recover our Born-Markov results:
\begin{eqnarray}\label{vindication}
\gamma(\vec n,\vec n') &=& 4\Gamma\sum_{\pm}\int\limits_0^\infty\!\! dz\,e^{-z^{2}} z^3\left( n_{\pm}n'_{\pm} +\mbox{sinc}^2\left( \frac{zs}{\lambda}\right) n_{\pm}n'_{\mp}\right)\nonumber\\
&=& 2\Gamma [(n_{+}n'_{+}+n_{-}n'_{-})+R(s/\lambda)(n_{+}n'_{-}+n_{-}n'_{+})]\nonumber\\
&\equiv& \Gamma\left([1+R(s/\lambda)]NN' + [1-R(s/\lambda)]nn'\right)\;,
\end{eqnarray}

where the dimensionless function $R$ is the same one defined in (\ref{Rdef}).

\section{Buffer gas mean free path}
Our exact calculation for our model thus confirms the Born-Markov approximation in its assumed limit of weak interaction, in spite of physical arguments against the predicted slow saturation of the decoherence rate at large separations. It is in hindsight unsurprising that the Gaussian guess proved false, since it was a guess about non-equilibrium behavior based solely on equilibrium properties.

When a physical argument is contradicted by a theoretical model, however, the physical argument often turns out not to be utterly wrong, but rather to have been a misapplication of rules from some real regime to which the model does not belong. In such cases it is usually an even bigger mistake to suppose that the model's regime is universal. If the physical argument is based on real observations, then at some point it will probably become valid, and the model will fail.

In our problem here, the physical expectation was that a certain quantity that depended on a separation distance should approach its limit at infinite separation faster than by a power law. That general kind of behavior does indeed occur in many physical systems. We can therefore ask whether a more rapid saturation of decoherence rate, such as seemed physically plausible initially, might indeed occur in some regime beyond the one we have so far considered in our model. Is there in general a length scale which we have overlooked until now, a scale other than the thermal coherence length $\lambda$, beyond which decoherence rates actually do saturate abruptly?

Indeed there is. So far, we have assumed that the buffer gas particles never collide with each other. Since we have also nonetheless assumed an initial equilibrium state for the buffer gas, we have implicitly considered that collisions do occur in the gas, and that we have prepared the reservoir by waiting many times the average interval between buffer gas collisions; but by neglecting buffer gas collisions in our computed time evolution, we have implicitly assumed that the separation distance between our two test system trapping sites is much shorter than the mean free path in the buffer gas (\textit{i.e.} the average distance a buffer gas particle travels before it collides with another buffer gas particle). While mean free paths in dilute gases can be made very long in realistic experiments, in general one could also consider a limit in which our two test system sites were separated by many mean free paths. 

Exact treatment of an interacting gas is not feasible, but the effect of collisions within the buffer gas can be investigated by imposing a finite coherence lifetime for buffer gas particles \cite{kadanoff1994} in the derivation of the Born-Markov master equation. To see how this works, let us focus first on the terms in (\ref{eq:rhotime}) which will be most affected by the decay factor we introduce. When (\ref{eq:rhotime}) is all expanded, the term which will appear in the place of $R(s/\lambda)$ in (\ref{eq:me}) will be\begin{widetext}
\begin{eqnarray}\label{decaygamma1}
	\tilde{R}(s,\lambda,L) &=& \frac{\hbar}{m\pi}\frac{\lambda^4}{s^{2}}\mathrm{Re}\left[\int_{0}^{\infty}\!dk\,k\int_{0}^{\infty}\!dk'\,k'\, 
\frac{
	e^{	-\beta	\frac{\hbar^{2}k'^{2}}	{2m}		}
	}{Z}
\,\sin(ks)\sin(k's)\int_{0}^{t}\!dt'\, e^{-i\frac{\hbar(k^2-k'^2)}{2m}(t-t')} e^{-\gamma_{kk'}(t-t')}\right]\;.
\end{eqnarray}
The real part is taken because the imaginary part of this quantity provides only a perturbation to the system Hamiltonian, rather than decoherence, and is therefore ignored throughout this paper.

The usual Born-Markov procedure at this point is to assume $\gamma_{kk'}\to 0+$ but take $t\to\infty$. Here we will instead keep a small but finite collisional coherence loss rate for the dilute buffer gas \cite{kadanoff1994}:
\begin{eqnarray}\label{}
	\gamma_{kk'}= \frac{1}{2}n_{B}\sigma_{B}\frac{\hbar}{m}(k+k') \equiv \frac{\hbar}{2mL}(k+k')
\end{eqnarray}
where $\sigma_{B}$ is the cross section for collisions between buffer gas particles (assumed independent of the cross section $\sim a^{2}$ for collisions between buffer and test particles), and hence $L = 1/(n_{B}\sigma_{B})$ is the mean free path. The $k,k'$ dependence of $\gamma_{kk'}$ represents the standard Boltzmannian assumption that the rate at which any buffer gas particle hits others is proportional to its speed.

Evaluating (\ref{decaygamma1}) for large $t$ we obtain
\begin{eqnarray}\label{decaygamma2}
	\tilde{R}(s,\lambda,L) &=& \frac{2\lambda^4}{s^{2}}\int_{0}^{\infty}\!dk\int_{0}^{\infty}\!dk'\,\frac{k\,k'}{k+k'}\, 
\frac{
	e^{	-\lambda^{2}k'^{2}	}
	}{Z}
\,\Bigl(\cos[(k-k')s]-\cos[(k+k')s]\Bigr)\Delta_{L}(k-k')\;.
\end{eqnarray}\end{widetext}
The factor
\begin{eqnarray}\label{}
	\Delta_{L}(k-k')\equiv\frac{1}{\pi}\frac{L^{-1}}{(k-k')^{2}+L^{-2}}
\end{eqnarray}
is a representation of a delta function $\delta(k-k')$ as far as any components of the integrand are concerned that vary slowly with $(k-k')$ on the scale $L^{-1}$. Since the simple ideal gas canonical ensemble that we use for the buffer gas would be dubious if the mean free path $L$ were not much longer than the thermal length $\lambda$, we assume now that $\lambda\ll L$. This means that the only term in the integrand of (\ref{decaygamma2}) for which $\Delta_{L}(k-k')$ might \textit{not} simply be a delta function is the one containing $\cos(k-k')s$. 

If $s\ll L$ as well as $\lambda\ll L$, as assumed in previous Sections of this paper, then $\Delta_{L}(k-k')$ acts as a delta function for this $\cos(k-k')s$ term as well, and the whole of (\ref{decaygamma2}) reduces simply to (\ref{Rdef}), yielding $\tilde{R}\to R(s/\lambda)$, and thus providing slow $1/s^{2}$ decoherence rate saturation with separation distance $s$. 

If $s\lesssim L$, however, then we instead obtain for the $\cos(k-k')s$ term
\begin{eqnarray}\label{}
	\int_{0}^{\infty}\!dk\,\frac{k}{k+k'}\cos[(k-k')s]\Delta_{L}(k-k') = \frac{1}{2}e^{-\frac{s}{L}}
\end{eqnarray}
to leading order in $k'L$. This then implies
\begin{eqnarray}\label{Rlim2}
	\lim_{s\gg L\gg\lambda}\tilde{R}(s,\lambda,L) = \frac{1}{2}\frac{\lambda^{2}}{s^{2}}e^{-s/L},
\end{eqnarray}
an exponential saturation of the decoherence rate to the limit of two independent scatterers on the scale of the reservoir mean free path $L$, instead of the long-range $\lambda^{2}/s^{2}$ behavior of $R(s/\lambda)$ according to our simple Born-Markov result (\ref{Rlimit1}).

If we repeat the above analysis for any other terms in the master equation (\ref{eq:me}), we find that everything either does not depend on $s$ at all (and so represents the saturated limit of decoherence at infinite separation), or else already decays with $s$ on the shorter length scale $\lambda$ and is negligibly affected by further decay on the longer scale $L$. Hence (\ref{Rlim2}) really represents the only significant effect of a finite but long mean free path $L\gg\lambda$ in the buffer gas.

So: we began this paper with a physical expectation that there should be some threshold separation distance between our two trapping sites, beyond which the relative number decoherence rate should quickly saturate to its value at infinite separation. We now see that this expectation was actually correct, but that the relevant threshold separation distance is the mean free path in the buffer gas, rather than its thermal coherence length.

In the previous section of this paper we argued that the $R(s/\lambda)$-term in our master equation is caused by scattering processes of one reservoir particle colliding with both of the system wells, thereby inducing correlations. In the present section, we showed that such correlation building effects are weakened by intra-reservoir interactions, ultimately increasing the number-difference decoherence rate $\Gamma(1-R)$. The important message of this part of the paper thereby is, that a consistent description of decoherence does not only requires the careful non-equilibrium treatment of system-reservoir interactions but also needs to take into account certain non-trivial elements of the detailed reservoir dynamics.

\section{Conclusions}
In this paper we have compared different theories for the decoherence of superpositions of different distributions of system particles between two sites, due to collisions with a reservoir of buffer gas. In particular we focused on the way in which decoherence rates depended on the distance between the two sites at which the system particles could be located. All our theories have been based on an idealized model system, but our results remain relevant to real phenomena in quantum gases, and should also be qualitatively indicative for a broad range of decoherence processes.

Our most primitive theory was simply an appeal to the thermal coherence length of the buffer gas as the decisive length scale. With it came an expectation that if our two sites were separated by more than a few coherence lengths, then decoherence should occur very much as if the separation were infinite.

Our second theory was the standard Born-Markov formalism, eliminating the buffer gas reservoir. This yielded a master equation of Lindblad form which indeed described decoherence, but we found that the approach of some decoherence rates to their infinite-separation values was only an inverse square decay, and not the sharper exponential or even Gaussian saturation that we had anticipated. Within our Born-Markov derivation we were able to identify physical factors responsible for the long-range sensitivity of decoherence to separation distance. On the one hand we saw that decoherence is a time evolution, not an equilibrium property, and that averaging over initial and final phases did not reproduce all the cancellations seen in thermal averaging, because initial and final phases are correlated. On the other hand we noted that interference between collisions at the two sites tends to preserve some entanglement between the sites.

To resolve any doubt as to whether the Born-Markov theory was really accurate enough for such subtle issues, we developed a third, more rigorous theory, consisting of asymptotic late-time approximations to the exact time evolution of our model. This theory confirmed the Born-Markov result in the regime of weak scattering; since Born-Markov had always assumed that regime explicitly, this was a vindication. This more rigorous theory also offered a much more general result for buffer gas decoherence, even with strong scattering. This general result may be of use in future studies of collisional decoherence.

Our fourth theory then returned to the vindicated Born-Markov procedure, but added collisions between reservoir particles to the model; this closed the circle of our analysis, by showing that rapid saturation of decoherence rates with separation distance does indeed occur in general. Exponential saturation sets in when the site separation is more than a few reservoir mean free paths. This principle may perhaps be tested directly in quantum gas experiments. It also indicates that decoherence in general may be sensitive to information transfer within the reservoir, as well as between reservoir and system.


\begin{appendix}
\begin{widetext}

\section{Time evolution of scattered bath particles}\label{AppendixEvolution}

If there are no bound states, the set of functions (\ref{ScatteringAnsatz}) form a complete and orthogonal set of energy eigenstates for buffer gas particles interacting with tightly localized system atoms as a pair of scattering centers. In cases where bound states do exist, we can show that their role is negligible, as explained in Appendix B. 

Consequently, the evolution of any initial wave function $\psi(\vec{r},0)$ of a bath particle is given by $\psi(\vec{r},t)=\int d^3k \, c_{\vec{k}} \, \phi_{\vec{k}} (\vec{r})\,e^{-\frac{i}{\hbar}E_k t}$, with the Fourier coefficients $c_{\vec{k}} = \int d^3r \phi_{\vec{k}}^*(\vec{r})\, \psi(\vec{r},0)$ (For notational simplicity, from here on we will drop the $\vec{n}$-subscript of $\phi_{\vec{k},\vec{n}}(\vec{r})$ ). For our specific problem the initial wave functions, according to (\ref{eq:scatinitialcondition}), are plane waves. Simply to avoid cluttering our equations with many primes on $k$ variables, we will write $\vec{k}_{0}$ in this appendix for the wave number of the initial plane wave that is being evolved, and use unprimed $\vec{k}$ as our integration variable. The Fourier coefficients of those states can be computed in a straightforward manner to yield
\begin{eqnarray}\label{FourierScattering}
c_{\vec{k}} & = & \delta^3(\vec{k}-\vec{k}_0) +\epsilon (k,k_0)\, \sum_{\pm}\alpha_\pm^* e^{\pm i \vec{k}_0\cdot \vec{s}/2}\;,\\
\epsilon (k,k^\prime) &= &\frac{1}{(2\pi)^2}\left( \mathcal{P}\frac{2}{k^2-k^{\prime 2}} + i \pi \frac{\delta(k-k^{\prime})}{k^\prime} \right)\nonumber
\end{eqnarray}
where $\mathcal{P}$ means that the Cauchy principal value has to be taken when necessary.

Inserting (\ref{FourierScattering}) into (\ref{eq:eigensol}) and performing  the angular integrals we find
\begin{eqnarray}\label{EvolvedWave}
\psi_{\vec{k}_0}(\vec{r},t) & =&  e^{-i\frac{\hbar t}{2m}k_0^2} \phi_{\vec{k}_0} + \frac{1}{2}\sum_{\pm} \alpha_\pm^0 \frac{e^{ik_0r_\pm}}{r_\pm}  \, \frac{e^{-i\frac{\hbar t}{2m}k_0^2}}{(2\pi)^{3/2}} -\frac{1}{2}\sum_{\pm} \alpha_\pm^{0*} \frac{e^{-ik_0r_\pm}}{r_\pm}  \, \frac{e^{-i\frac{\hbar t}{2m}k_0^2}}{(2\pi)^{3/2}}  \nonumber\\
&&-\frac{1}{\pi i} \sum_{\pm} \mathcal{P}\int_{-\infty}^{+\infty} \frac{dk}{\Lambda(k)}\, \frac{k a_\pm}{k^2-k_0^2}\, \frac{e^{-i\frac{\hbar t}{2m}k^2}}{(2\pi)^{3/2}}\, \frac{e^{ikr_\pm}}{r_\pm} \left( (1+ika_\mp) e^{\pm i\vec{k}_0 \cdot \vec{s}/2} - a_\mp\frac{e^{iks}}{s} e^{\mp i\vec{k}_0 \cdot \vec{s}/2}\right)
\end{eqnarray}
where we denoted $\alpha_\pm^0 = \alpha_\pm(\vec{k}_0)$ and $\Lambda(k) = (1+ika_+)(1+ika_-) - \frac{a_+a_-}{s^2}e^{2iks}$. Here we have used the identity 
\begin{eqnarray}\label{Ortho}
 \frac{k+k^\prime}{2}\left( \alpha_+^* \alpha_+^\prime + \alpha_-^* \alpha_-^\prime \right) 
+ \frac{e^{i k^\prime s}-e^{-i k s}}{2i s}\left( \alpha_+^* \alpha_-^\prime \right. &+&\left. \alpha_-^* \alpha_+^\prime \right) =
\nonumber \\
= \frac{i}{2}\left( \alpha_+^\prime e^{-i \vec{k} \vec{s}/2}  - \alpha_+^* e^{i \vec{k}^\prime  \vec{s}/2} 
 \right. &+& \left.\alpha_-^\prime e^{i \vec{k}\vec{s}/2} - \alpha_-^* e^{-i \vec{k}^\prime \vec{s}/2} \right)\;,
\end{eqnarray}
\end{widetext}
which is obtained by writing out the orthogonality condition, collecting products of the scattering coefficients $\alpha_{\pm}$, and using the given $\pm k$-symmetry to extend the lower integral bound to $-\infty$.

\begin{figure}
\begin{center}
\includegraphics[width=0.5\textwidth]{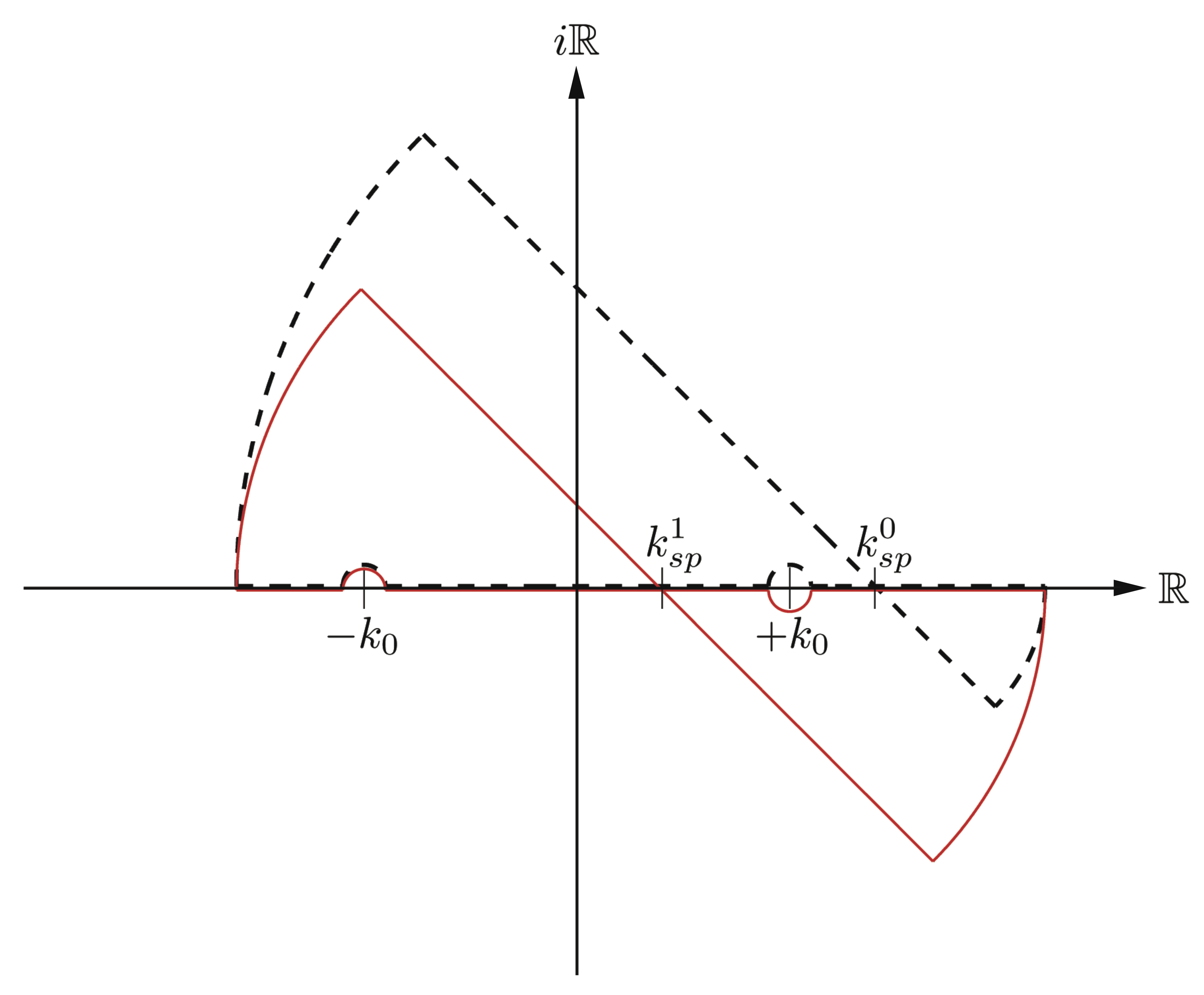}
\end{center}
\caption{Contour of integration for the $k$-integral in (\ref{EvolvedWave}), for some arbitrary fixed $r_{\pm}$, at a time before (black dashed line) and after (red solid line) the time-dependent saddle point $k_{\mbox{sp}}=\frac{m}{\hbar t}r_{\pm}$ reaches the pole in $+k_0$. The points $k_{\mbox{sp}}^{0,1}$ here denote the saddle points at the two different times.}
\label{fig:Contour}
\end{figure}

We consider such a dilute buffer gas that decoherence only becomes significant over long time scales, and thus we are only interested in those effects of the reservoir which grow secularly in time. We must therefore focus on the long time limit of the reservoir particles' evolution, discarding transient effects. For such large times the integral written out in the second line of equation (\ref{EvolvedWave}) can be evaluated by the method of steepest descents. To evaluate (\ref{EvolvedWave}) for large times $t$, we analytically continue the integrand of (\ref{EvolvedWave}) into the complex $k$-plane and use the closed contour shown in Fig.~\ref{fig:Contour} to determine the integral over the real axis. 

The closed path integral consists of four parts: the principal value over the real axis (which is the contour along which we actually need to integrate), the two small arcs around the poles in $k=\pm k_0$, the two large arcs at $k=\pm\infty$ ranging into the complex plane and the integral along the path of stationary phase $k-k_{\mbox{sp}}=\xi e^{-i\pi/4}$, which crosses the real k-axis at the saddle point of the exponent $k_{\mbox{sp}}=\frac{m}{\hbar t}r_\pm$. We will choose the path such that no poles are contained inside the enclosed area so that, according to Cauchy's residue theorem, all four contributions add to zero. But, because the position of $k_{\mbox{sp}}$ is time- and position-dependent, the integral path circumvents the pole at $+k_0$ either clockwise ($k_{\mbox{sp}}<k_0$) or anti-clockwise ($k_{\mbox{sp}}>k_0$). Therefore, the arc around the $+k_0$-pole will yield contributions whose sign will depend on $t$ and $r_{\pm}$. These contributions turn out to describe the propagation of the outgoing scattered wavefront.\\

We will now address each part of the integral path shown in Figure \ref{fig:Contour} separately. Since the integrand is exponentially small for any points lying on the circular arcs at $|k|=\infty$, their contribution vanishes. 

The arcs around each of the two poles yield plus or minus half of the residue, depending on the direction of integration. As $k_{\mbox{sp}}>0$, the contribution from the pole in $-k_0$ will always cancel the incoming s-wave of the third term in (\ref{EvolvedWave}). Contrarily, the sign of the contribution of the other pole at $k=+k_0$ will depend on whether the saddle point is on the right $k_{\mbox{sp}}>k_0$ or on the left $k_{\mbox{sp}}<k_0$ of the pole: in the first case, its contribution yields exactly the same value as the outgoing s-wave written out in the second term of (\ref{EvolvedWave}). Together they cancel the outgoing s-wave of $\phi_{\vec{k}_0}$ in the first term of (\ref{EvolvedWave}), leaving only a plane wave. In the second case the pole contribution and the second term of (\ref{EvolvedWave}) cancel, and a spherical wave remains besides the plane wave. Consequently, the full solution will have an outgoing s-wave for $\frac{\hbar k_0}{m}t > r_\pm$, and no such contribution for $\frac{\hbar k_0}{m}t < r_\pm$. This abrupt time- and position-dependence of one of the s-wave contributions describes the expansion of a spherical wavefront, when the smoother profile of the exact wavefront is neglected as a transient effect (because no effects of this smooth wavefront will grow secularly with $t$). The steady expansion of the wavefront, neglecting its transient edge, may thus be expressed in terms of a unit-step function $\Theta\left( \frac{\hbar k_0}{m}t - r_\pm\right)$ in front of the spherical wave terms.\\


As a last step we may evaluate the integral along the path of steepest descent, where the integrand consists, for almost all $r,t$, of a slowly varying function in $k$ multiplied by a sharp Gaussian around $k=k_{\mbox{sp}}$. To a very good approximation, the integral is therefore given by the integral of the Gaussian times the remaining integrand evaluated at the saddlepoint $k=k_{\mbox{sp}}$:

\begin{eqnarray}\label{SPcontribution}
\sqrt{2\pi}\left(\frac{m}{\hbar t} \right)^{3/2} e^{i \frac{m}{2\hbar t}r_\pm^2}\,\left[\frac{\bar\alpha_\pm}{k^2-k_0^2} \right]_{k_{\mbox{sp}}}\nonumber\\
\bar\alpha(k)_\pm = a_\pm\frac{ A_\mp(k) e^{\pm i\vec{k}_0\cdot\vec{s}/2} -B_\mp(k) e^{\mp i\vec{k}_0\cdot\vec{s}/2}}{\Lambda(k)}\;.
\end{eqnarray}

\begin{figure}
\begin{center}
\includegraphics[width=0.5\textwidth]{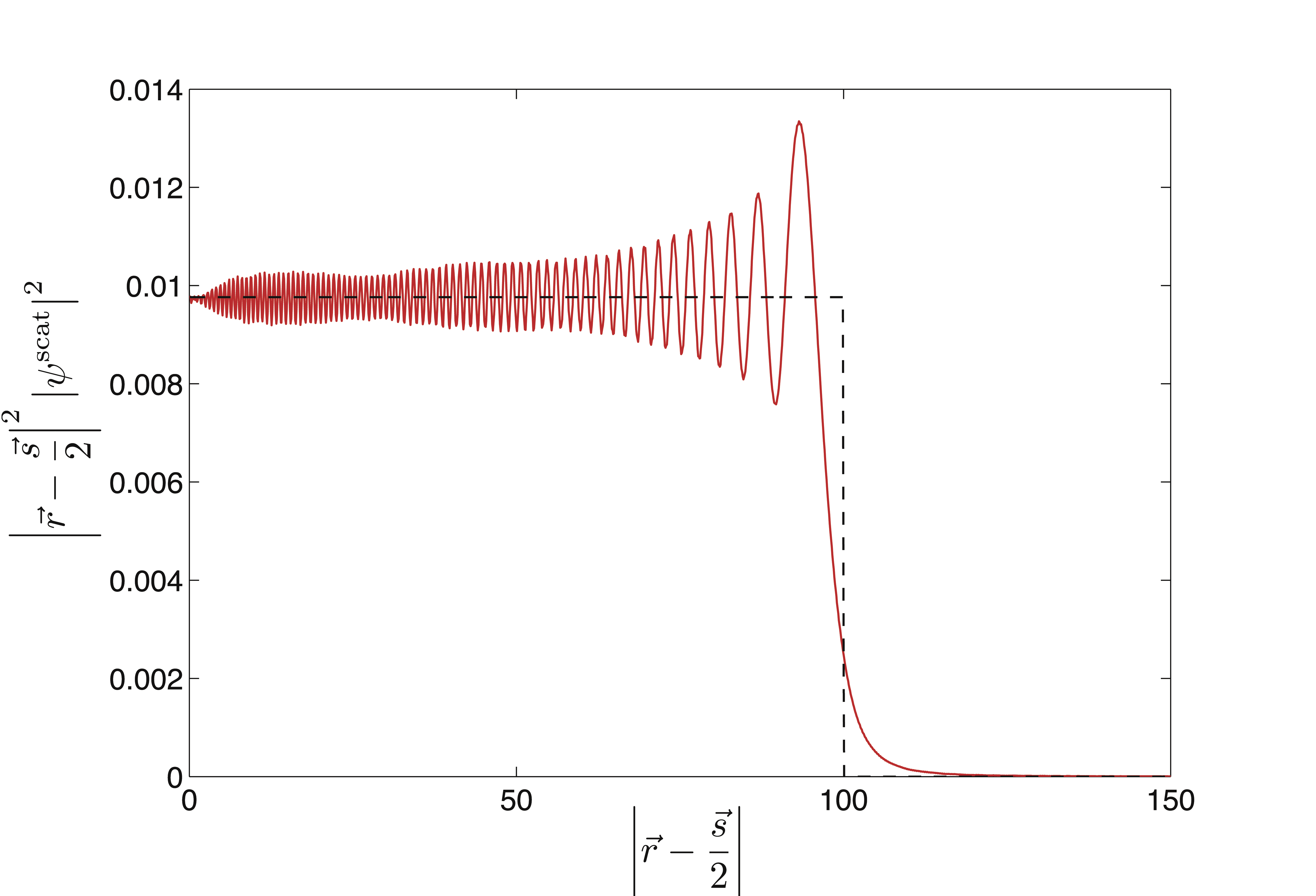}
\end{center}
\caption{Comparison between the outgoing spherical wave from one of the scatterers obtained by steepest descent (black dashed line) and by numerical integration in (\ref{EvolvedWave}) (red solid line) for $a_\pm = 3/2,1/2$, $k_0=10$ and incidence angle $\pi/6$ at time $t=10$ (the separation $s$ has been taken as length unit).}
\label{fig:SPcomparing}
\end{figure}

Except when $k_{\mbox{sp}}$ is close to $k_0$, the factor $\left[\frac{\bar\alpha_\pm}{k^2-k_0^2} \right]_{k_{\mbox{sp}}}$ is slowly varying and bounded, and the integral along the saddle point path decays at long times as $t^{-3/2}$.  This means that this path's contribution to the time-dependent wave function is only significant for $r$ close to $\frac{\hbar  k_0}{m} t =r_\pm$, \textit{i.e.} within the transient layer of the spherical wavefront. Numerical integration confirms that the contributions are indeed significant in this region (see Figure \ref{fig:SPcomparing}), but that their contribution to any total probability remains bounded and does not grow steadily with time as the scattered wave expands.


Neglecting the steepest descent path transient contribution, all other leading order terms together can be written as
\begin{eqnarray}
\psi(t) & = & \psi_0 - \psi^{\mathrm{scat}} \nonumber\\
\psi_0(t) & = &  \frac{e^{-i\frac{\hbar t}{2m}k_0^2}}{(2\pi)^{3/2}}\,e^{i\vec{k}_0 \cdot \vec{r}}\nonumber \\
\psi^{\mathrm{scat}}(t) & = & \frac{e^{-i\frac{\hbar t}{2m}k_0^2}}{(2\pi)^{3/2}}\, \sum_{\pm} \Theta\left( \frac{\hbar t k_0}{m} -r_\pm\right)\, \frac{e^{ik_0r_\pm}}{r_\pm} \,\alpha_\pm^0  \nonumber
\end{eqnarray}
where the  step function, resulting from having the saddle point either on the right or on the left of the pole at $+k_0$, describes an outgoing spherical wavefront. 

We emphasize again that the sharp wavefront described by the step functions here is an approximation that neglects transients, but captures all effects that grow secularly with $t$. An exact treatment, without any steepest descent approximation, will of course yield a softer wavefront, which moreover includes oscillations, as shown in Fig.~\ref{fig:SPcomparing}. We see, however, that the unit step function correctly describes the envelope of the numerical result for the spherical waves $\psi^{\mathrm{scat}}$. It is only this envelope which produces the steady growth in scattering probability that gives the constant decoherence rates at late times. The questions in this paper concern only those constant rates, and so neglecting transients in our time evolution, and retaining only the sharp-fronted envelope functions, provides the exact answers to the questions we are discussing.

\begin{widetext}
\section{Completeness of the set of energy eigenstates}\label{AppendixBound}

In the preceding appendix we calculated the time evolution of a plane wave, assuming that the set of scattering states proposed in Ansatz (\ref{ScatteringAnsatz}) is complete and orthogonal. We will show here that this is true, except when bound states exist. We will then further determine the conditions for the existence of bound states, find the bound state wave functions explicitly, and explain why their contribution to constant asymptotic decoherence rates at late times is exactly zero in any case. This establishes the general validity of our previous appendix, regardless of the presence of bound states.\\

Testing our set of eigenfunctions for completeness, we find that the completeness relation evaluated with the set of solutions (\ref{ScatteringAnsatz}) does not always yield the pure delta function $\delta(\vec{r}-\vec{r}\,^\prime)$ that it should, but in general contains an additional integral:
\begin{eqnarray}\label{eq:phicompleteness}
 \int d^3k\, \phi_{\vec{k}}^*(\vec{r}\,)\,\phi_{\vec{k}}(\vec{r}\,^\prime) &=& \delta^3(\vec{r}-\vec{r}\,^\prime) -
 2\pi i \sum_{\pm}\int_{-\infty}^{+\infty} dk\, k \, \left(  \frac{a_+ a_-}{\Lambda} \,
 \frac{e^{iks}}{s} \frac{e^{ikr_\pm}}{r_\pm}\frac{e^{ikr^\prime_\mp}}{r^\prime_\mp} 
 \frac{a_\pm (1+i k a_\mp)}{\Lambda}\frac{e^{ikr_\pm}}{r_\pm}\frac{e^{ikr^\prime_\pm}}
	{r^\prime_\pm}\right)\;.
\end{eqnarray}
\end{widetext}
This additional integral may be identically zero, but if the scattering amplitudes have poles in the upper half of the complex plane, then it is not. Such poles in the scattering amplitudes correspond to bound states; this may be seen by seeking additional bound state wave function solutions with negative energies $E_n = -\frac{\hbar^2 q_n^2}{2m}$ and wavefunctions of the form
\begin{equation}\label{BoundstateWaveFunction}
\phi_n(\vec{r}) = \sum_{\pm} w_{\pm}^n\,\frac{e^{-q_n r_\pm}}{r_\pm}
\end{equation}

Inserting (\ref{BoundstateWaveFunction}) in the time-independent Schr\"odinger equation, one finds that the condition for the existence of nontrivial solutions is
\begin{equation}\label{eq:qncondition}
\left(q_n s - \frac{s}{a_-}\right)\left( q_n s - \frac{s}{a_+}\right) = e^{-2q_n s}
\end{equation}
which is exactly the condition for the scattering amplitudes to have poles on the positive imaginary axis: $\Lambda(k_n)=0$, at points $k_n=iq_n$.

\begin{figure}
\begin{center}
\includegraphics[width=0.45\textwidth]{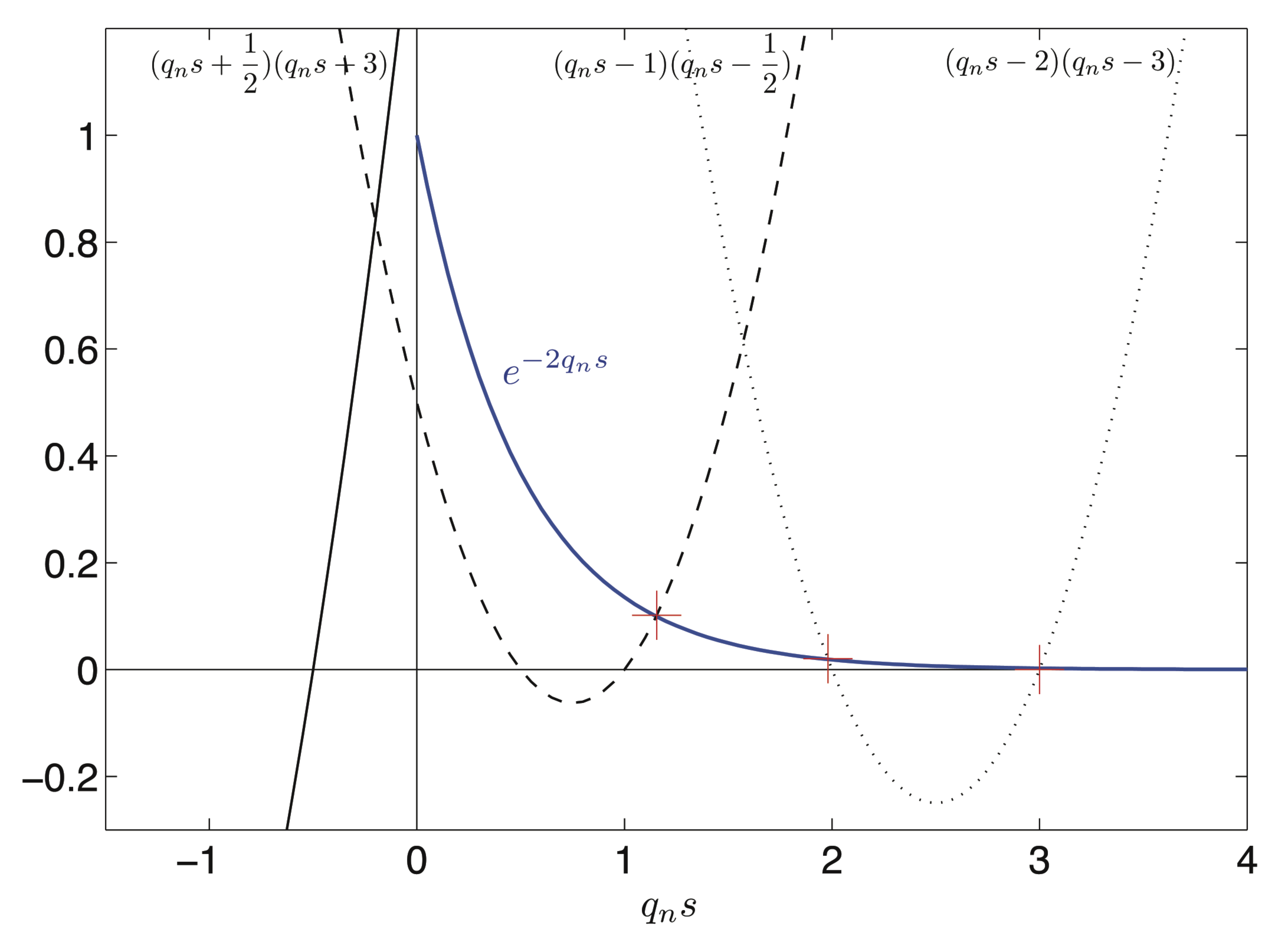}
\end{center}
\caption{Examples of crossing points of curves $e^{-2q_n s}$ (blue line) and $\left(q_n s - \frac{s}{a_-}\right)\left( q_n s - \frac{s}{a_+}\right)$ (black lines) for different values of $\frac{a_\pm}{s}$. The crossing points (red) determine the roots of $\Lambda(k)$, i.e. energies of bound states when there are two (dotted line, with $\frac{s}{a_\pm}=2, 3$), one (dashed line, $\frac{s}{a_\pm}=1, \frac{1}{2}$) or no such bound states (solid line, $\frac{s}{a_\pm}=-3, -\frac{1}{2}$).}\label{fig:BSroots}
\end{figure}

As Fig.~\ref{fig:BSroots} illustrates, the equation $\Lambda(iq_n)=0$ has two positive roots when $a_{\pm}>0$ and $a_+ a_-<s^2$; it has no positive roots when $a_{\pm}<0$ and $a_+ a_-<s^2$; and otherwise it has exactly one positive root. Therefore, depending on the separation between wells and the s-wave scattering lengths $a_{\pm}$, bound states can exist. When they exist, they must be included as discrete terms along with the integral over the continuum of scattering eigenstates, in order to have a complete set of states.

For the set of $q_n$'s which fulfill (\ref{eq:qncondition}), we can further determine the coefficients 
\begin{equation}
w_\pm^n= \frac{\chi_n}{\sqrt{2\pi}}\sqrt{q_n a_\pm (1-q_na_\mp)}
\end{equation}
where $ \chi_n$ is related to the residue of the scattering amplitudes in the pole $iq_n$
\begin{eqnarray}\label{ResidueLambda}
\chi_n^2 &=&  \frac{s}{s(a_++a_-)+a_+a_-(q_ns-2e^{-2q_ns})}\nonumber\\
i\chi^2_n &=&\mbox{Res}\left\{ \Lambda^{-1}(k)\right\}|_{k=iq_n}
\end{eqnarray}\\
Including these states in the sum over eigenstates ensures completeness of the set $\{\phi_n(\vec{r}),\phi_{\vec{k}}(\vec{r})\}$, as the integral in (\ref{eq:phicompleteness}) is exactly canceled by the additional contributions.

The inclusion of bound states thus leads to additional terms in the scattering time evolution, so that
\begin{eqnarray}
\psi(\vec{r},t)&=&\int d^3k \, c_{\vec{k}} \, \phi_{\vec{k}} (\vec{r})\,e^{-\frac{i}{\hbar}E_k t}
+ \sum_n c_n\, \phi_{n}(\vec{r})\, e^{-iE_nt/\hbar}\;,\nonumber\\
c_n &=& \frac{1}{\sqrt{2\pi}}\sum_\pm \frac{w_\pm^n}{q_n^2+k_0^2}\, e^{\pm i \vec{k}_0 \cdot \vec{s}/2}\nonumber\;.
\end{eqnarray}

In contrast to the unbound continuum solutions, however, for which the support of the scattering wave function grows linearly in time, 
\begin{equation}
\int d^3r\,|\psi^{\mathrm{scat}}(\vec{r},t)|^2 \propto \sum_{\pm}\int \frac{d^3r}{r_\pm}\, \Theta\left( \frac{\hbar k_0}{m}t - r_\pm\right) \propto t\;,\nonumber
\end{equation}
any bound state wave function has finite norm and decays linearly in space. Hence, the total probability carried by these states remains fixed at its initial value $|c_{n}|^{2}$.
Consequently, in the long time limit, any bound state contribution is of order $t^{0}$ when compared to the secularly growing contribution $\propto t^{1}$ from the continuum of unbound solutions. For the purpose of computing asymptotically constant decoherence rates at late times, which is the only purpose of this paper, the existence of bound states in the eigenstate spectrum of the two-scatterer potential can be ignored without loss of generality.

\end{appendix}

\bibliography{bib}

\end{document}